\documentclass[a4paper,fleqn,usenatbib]{mnras}

\usepackage[T1]{fontenc}
\usepackage{ae,aecompl}

\usepackage{savesym}
\usepackage{graphicx}	
\usepackage{amsmath}	
\usepackage{pdflscape}
\usepackage{adjustbox}
\savesymbol{iint}
\usepackage{amssymb}
\usepackage{txfonts}
\restoresymbol{TXF}{iint}

\usepackage{multicol}        
\usepackage{bm}	
\usepackage{rotating}
\title{The relations between dust properties and galaxy global / integrated quantities in the nearby Universe}  
\author[B. A. Pastrav]{Bogdan A. Pastrav$^{1}$\thanks{E-mail:bapastrav@spacescience.ro}
\\
$^{1}$Astrophysics, Cosmology and Theoretical Physics Laboratory, Institute of Space Science-INFLPR Subsidiary, Atomistilor 409, 077125, Bucharest-Magurele, Romania\\
}
\date{Accepted xxxx xxxxxx xx. Received xx xxxxxxx xx; in original form xxxx xxxxxxx xx.}

\pubyear{2025}

\begin{document}
\label{firstpage}
\pagerange{\pageref{firstpage}--\pageref{lastpage}}
\maketitle

\begin{abstract}

Results of a case study of a sample of low-redshift galaxies are presented, to determine dust temperatures and emissivity indices through a less time-consuming
method, and to connect both global and integrated galaxy properties with those of dust, ISM and star-formation. Dust temperatures ($T_{d}$) are determined based on the 
corresponding galaxy dust masses, independently calculated in our previous work, through a self-consistent method, without the need to actually perform a complete spectral
energy distribution (SED) fit of the cold dust emission fluxes. The range and average dust temperatures are found to be consistent within errors with values from other studies. 
Simultaneously, the dust emissivity indices ($\beta_{d}$) are determined, and their evolution with temperature quantified, with the $T_{d}$ anti-correlation still being present. It is investigated whether $\beta_{d}$ can be
predicted from other relation or if it scales with other integrated dust / ISM or galaxy property, which could be used as a proxy. In this respect, new and established relations between $T_{d}$, $\beta_{d}$, the dust surface density and global / integrated
galaxy and star-formation related quantities are presented and analysed. We find that SFR, sSFR or $\Sigma_{SFR}$ are inconclusive traces of the dust temperature.  We also find that the extent of dust emission distribution is slightly lower on average, but
comparable with the optical stellar continuum emission one. The results and conclusions can be relevant for larger scales studies of low to mid-redshift galaxies from the
latest surveys.

\end{abstract}

\maketitle

\section{Introduction}\label{sec:intro}

Dust has a central role in galaxy evolution and star-formation, regulating many processes in the interstellar medium (ISM) of galaxies, as it absorbs and scatters more than
one third up to half of the ultraviolet (UV) radiation coming from young and old stars alike (\citealt{Dri07, Bendo14, Bia18}), and radiates it at longer wavelengths - near-infrared
(NIR) to sub-milimetre (\citealt{Draine03, Gal18}). It is a catalyst for molecular hydrogen formation and, as dust grains are composed mainly of metals, has a direct influence
on the metallicities of galaxies. Dust grains of different sizes show different ranges in temperatures (\citealt{LiDraine01}) - small grains suffer stochasting heating from
star-forming regions, with huge variations in temperature (to very high values, from 100K up to 1000K), while large grains are in a state of relative thermal equilibrium 
with the interstellar radiation field (having temperatures in the range 10-35K), emit mostly in the infrared, and constitute the dominant component of the total dust mass
within galaxies (\citealt{Vla05}) and of the FIR total flux (\citealt{DraineLi07}).

Determining the dust properties - mainly the dust mass, dust temperature and its emissivity index, can be a challenging task if not enough NIR to far-infrared
(FIR) and sub-mm photometric data are available to fully sample the spectral energy distribution of the dust. These properties are pivotal in studies
of star-formation (to determine how much dust is obscuring the star-formation regions) and interstellar medium evolution. The dust emissivity index (also known as dust
mass absorbtion coefficient) is essential when deriving dust masses from spectral energy distribution fits of the cold dust emission, but it is also a parameter tightly 
connected to the properties of dust grains, such as their composition (the proportion of silicates vs graphite) and structure (crystalline or amorphous), and the dust
temperature. In most theoretical models the dust emissivity index has been fixed to a value $\beta_{d}=2$ (e.g. \citealt{Wei01, Draine03}), but from observations its value
was not accurately and reliably constrained as one needs available data in the Rayleigh-Jeans part of the cold dust SED, and therefore available sub-mm and milimetre data.
The average observational values measured for nearby galaxies and the Milky Way range between 0.8 and 2.5 (\citealt{YangPhil07, Para10, Juv11, Boss12, Gala12, Smi12, Cle13,
Remy13, Cor14, Gala14, Planck14, Grossi15, Planck15, Taba14, Lamp19, Bendo24}).\\
To determine the dust mass, temperature and its mass absorption coefficient, a routinely used single-temperature modified black-body (MBB) has been proven to describe quite
well the cold dust SED at wavelengths above \textasciitilde100$\mu m$, for regions in our own Galaxy, in nearby Universe but also for high-redshift galaxies (e.g. \citealt{Para10,
Ski11, Bia13, Pozzi20, Tang21, Wit22, Alg24}), for most cases where the optically thin aproximation is valid (\citealt{EaEdm96}). To extract these three parameters from
SED fits for large samples is computationally expensive and in many studies, sometimes the dust mass absorbtion coefficient was constrained to a value $\beta_{d}=2$ (e.g. 
\citealt{Dunne00, Vla05, Ski11, Hunt15}). However, it was been observed that a single temperature for the whole dust distribution is not adequate for all galaxies, and therefore
two MBB were used, one for each of warm and cold dust components (\citealt{DunEal01, Vla05, Gala12, Gala14, Kirk14, Taba14}), resulting in dust masses approximately twice higher 
than before. As variations in $\beta_{d}$ between and within galaxies were observed, in other works it was left as a free parameter in the SED fitting procedure 
(\citealt{Boss12, Gala12, Smi12, Cle13, Remy13, Cor14, Taba14, Gala14, Kirk14, Grossi15, Lamp19, Bendo24}). Still, the use of a single MBB is a simplification of dust emission treatment, and the output parameters
can be looked as apparent or effective temperatures and emissivity indices, as there is a mix of warm and dust components along the line of sight, with various size 
distributions, grain compositions and therefore, grain emissivities. The dust masses however, are more robust quantities, not so much affected by the use of free or fixed
$\beta_{d}$ (\citealt{Cor14}). Nevertheless, this method provides meaningful results and can help us investigate and understand the variations of dust properties with other
galaxy and star-formation relevant global parameters.

This is the fourth paper of the series, following from \cite{Pas20} (Paper I, where we focused on dust effects on disc scaling relations), \cite{Pas21} (Paper II, where
the main bulge and early-type galaxy scaling relations where analysed, together with black-hole scaling relations and criteria for bulge and galaxy classification) and 
\cite{Pas24} (Paper III, where we concentrated on the star-formation and dust/ISM scaling relations and deriving more instantaneous dust-corrected star-formation rates 
(SFR) using unattenuated $H\alpha$ luminosities).\\
Here, the goal of this case study is to: a) derive accurate dust temperatures and corresponding dust emissivity indices for the same galaxies analysed in Papers I and III (for
consistency and validation of the method), using previous work prescriptions and the dust masses independently (of this study) derived in Paper I, through a self-consistent
approach; b) investigate the $T_{d}-\beta_{d}$ evolution and its potential degeneracy; c) check for possible tracers for the dust temperature among the star-formation 
related quantities, that could be used to estimate $T_{d}$ and $\beta_{d}$ when there are not enough photometric data available to extract these from a spectral energy 
distribution (SED) fit; d) investigate / analyse the relations - new or confirmed - between $\Sigma_{d}, \beta_{d}$, metallicity ($Z[O/H]$), SFR, their implications for 
star-formation and ISM evolution, and which correlations are tighter and fundamental. A widely used modified blackbody function (MBB) - dependent on $T_{d}$, $\beta_{d}$ and
frecquency ($\nu$), is considered to characterise the cold dust emission SED in the NIR-submm wavelength range. While for this single temperature dust emission SED usually multiple fluxes / 
luminosities are needed at various wavelengths in the specified range (and sometimes not available) to derive the dust masses through SED fitting (and subsequently the dust
temperatures and emissivity indices), through this present approach only the flux / luminosity at 250$\mu m$, characteristic for the peak of the cold dust emission, is 
required, while $\beta_{d}$ is left as a free parameter. Thus, the whole cold dust emission SED will not be fitted (less computationally expensive) because the dust masses for
this sample are apriori known (from Paper I, or could be calculated for other samples of galaxies using the method presented in the same paper) and independently derived. 
For some of the conclusive corrected relations we investigate the degree of correlation between the parameters (where relevant), calculate the scatter of these relations 
and analyse the implications of the main results for ISM evolution and star-formation. We then discuss these results and compare with other relavant studies on low-redshift
galaxies.\\
The procedure described in this paper can be used in conjunction with the method to derive dust masses from Paper I (and potentially the prescriptions to derive SFR from 
Paper III) in larger scale studies of ISM evolution and star-formation, if needed and suitable optical+FIR photometry data are available.

The paper is organised as follows. In Sect.~\ref{sec:sample} we present briefly the galaxy sample used in this study, while in Sect.~\ref{sec:method} we describe the method 
used for the determination of dust temperature and dust emissivity index. In Sect.~\ref{sec:results} we present the main results - the relations between $T_{dust}, \beta_{d}$,
the surface density of dust, the star-formation or stellar mass, and other related global and integrated quantities, together with all their characteristic parameters. In 
Sect.~\ref{subsec:Rdust_vs_Rdisk_RHalpha} we show how the extent of dust distribution compares with the optical stellar emisssion one. The results are discussed in relation
with other relavant studies in the literature. In Sect.~\ref{sec:discussion} we discuss upon the possible sources of errors, differences with other studies, and the limitations
of the method, while in Sect.~\ref{sec:conclusions} we summarise the important results obtained in this study and draw some conclusions.

\section{Sample}\label{sec:sample}

Our sample of 24 nearby galaxies - the same as in \cite{Pas24}, is used as case study for consistency of the analysis and validation of the method. It comprises 19
low-redshift spiral galaxies and 5 lenticulars, included in the SINGS (\textit{Spitzer} Infrared Nearby Galaxies Survey; \citealt{Ken03}) survey and the KINGFISH project 
(Key Insights on Nearby Galaxies: a Far-Infrared Survey with \textit{Herschel}; \citealt{Ken11}). The galaxies were already analysed in B band in Paper I and Paper II, 
while in Paper III the images of the galaxies were analysed at the H$\alpha$ line wavelength. As before, we did not considered barred, dwarf and irregular galaxies from
the KINGFISH sample, because we want to observe dust-free scaling relations, and at this point we cannot properly account for the effects of dust on the photometric and
structural parameters of the former (barred galaxies), or for the more peculiar geometry of the latter (dwarfs and irregulars). Ellipticals from the KINGFISH survey are 
also not considered as we focus here on more dusty galaxies to derive the dust characteristics ($T_{d}$ and $\beta_{d}$) and their relation with star-formation and dust/ISM
relevant quantities. \\
More details about the galaxy images and their characteristics are given in Papers I, II and III, and therefore we do not repeat them here.
The KINGFISH project is an imaging and spectroscopic survey, consisting of 61 nearby (d<30 Mpc) galaxies, chosen to cover a wide range of galaxy properties (morphologies,
luminosities, SFR, etc.) and local ISM environments typical for the nearby Universe, being therefore representative for the population of typical low redshift galaxies.

\section{Method}\label{sec:method}

Dust temperatures are determined based on the galaxy dust masses previously calculated in \cite{Pas20} through a self-consistent method. The details are given below. The
star-formation rates and related quantites (specific star-formation rates - sSFR, SFR surface density - $\Sigma_{SFR}$), both observed and intrinsic, which appear in some
of the investigated relations in this study, have also been determined previously in Paper III (see Table 4 for numerical values and Section 3 for the method) from 
integrated H$\alpha$ luminosities and previous work prescriptions and relations.

\subsection{Deriving dust temperature and dust emissivity index}\label{subsec:dust_temp_beta}

For a better understanding, we first remind the reader very briefly what relations were used to derive the dust masses for the whole sample, $M_{d}$, in Paper I (see there
section 3.3). Prior to deriving dust masses, the empirical correlation between central face-on dust opacity in the B band ($\tau_{B}^{f}$) and the stellar mass surface density
($\mu_{*}$) of nearby spiral galaxies, found by \cite{Gro13} (Eq.~\ref{eq:Grootes}), was used to determine the $\tau_{B}^{f}$ values. This quantity was essential when 
applying the corrections for dust effects for all photometric and structural parameters, but also for the subsequent determination of dust masses, using Eqs. (2) and 
(A1-A5) from \cite{Gro13}. We show this below in Eq.~\ref{eq:Mdust_tauB}. This relation is based on the fixed large scale star-dust geometry of the \citealt{Pop11} model,
where the diffuse dust in the disk (which mostly determines the optical depth of a spiral galaxy) is considered to have an axisymmetric distribution, as two exponential 
discs (see also Eq. (44) in \citealt{Pop11}).
\begin{eqnarray}\label{eq:Grootes}
\log(\tau_{B}^{f})=1.12(\pm0.11)\cdot\log(\mu_{*}/M_{\odot}kpc^{-2})-8.6(\pm0.8) 
\end{eqnarray}

\begin{eqnarray}\label{eq:Mdust_tauB}
\tau_{B}^{f}=K(B)\frac{M_{d}}{R_{s,d}^{2}(B)}  
\end{eqnarray}
where $K(B)$ is a constant containing the details of the dust geometry and dust spectral emissivity ($k_{\nu}$) of the \cite{Draine03} model at the B band wavelength,
while $R_{s,d}(B)$ represents the observed B band scale-length of the stellar disc (obtained in Paper I).  All the values of $M_{d}, R_{s,d}(B), \tau_{B}^{f}$ and $\mu_{*}$ are
tabulated in Tables 2 \& 3 from Paper I.

To characterise the dust emission spectral energy distribution in the NIR-submm range (above 70-100$\mu m$) - the cold dust emission SED, and considering the dust grains
to be in local thermal equilibrium, we use a modified blackbody function as in the following relation. This is because it was shown in many studies (see the references in 
Sec.~\ref{sec:intro}) that it approximates quite well this part of a galaxy SED, in the optically thin limit. It is still an oversimplification does not take into account
that dust within a galaxy most likely has a range of dust temperatures (different dust heating sources, e.g. \citealt{Bendo14}) and therefore multiple dust components with 
different grain sizes along the line of sight. Nevertheless, meaningful conclusions about the overall dust temperature and dust grain sizes (through the derived $\beta_{d}$
value) can be extracted from such fits. We explain in Sect.~\ref{sec:discussion} why using two MBB functions (one for each of the warm and cold dust components) is not feasable
in this study, with the present approach.
\begin{eqnarray}\label{eq:MBB}
F_{\nu}=\frac{M_{d}}{d_{gal}^2}\kappa_{\nu}B(\nu,T_{d})=\frac{M_{d}}{d_{gal}^2}\kappa_{\nu_{0}}(\frac{\nu}{\nu_{0}})^{\beta_{d}}B(\nu,T_{d})  ,
\end{eqnarray}
where $F_{\nu}$ is the observed flux density at the frecquency $\nu$, $d_{gal}$ is the distance to the galaxy, $\kappa_{\nu_{0}}$ represents the dust mass absorbtion 
coefficient at the frecquency $\nu_{0}$, while $\beta_{d}$ is the dust spectral emissivity index (it gives the variation of the dust mass absorbtion coefficient with 
$\nu$; \citealt{EaEdm96})
\begin{eqnarray}\label{eq:kappa_nu}
 \kappa_{\nu}=\kappa_{\nu_{0}}(\frac{\nu}{\nu_{0}})^{\beta_{d}}   
\end{eqnarray}
For this study, our choice is to use the flux density at $\lambda=250\mu m$, a data point far enough in the IR on the SED (and therefore not affected by dust) which constrains
/ characterises well the peak of the cold dust emission and is dominated by the thermal emission of the cold dust. The flux density at 250$\mu m$ was also used in \cite{Dale12} 
and \cite{Gro13} to derive dust masses, while \cite{Gala12} and \cite{Ski11} considered the flux densities at $\lambda=500\mu m$, which is known to be less sensitive at 
variations of the dust temperature. For $\nu_{0}$ we use the value corresponding to
$\lambda_{0}=100\mu m$, while for $\kappa_{\nu_{0}}$ we take the value at $\lambda_{0}$ from \cite{Draine03}. As pointed out by \cite{Bia13}, using a $\kappa_{\nu_{0}}$ 
value derived from a dust model that uses a certain fixed value for dust spectral emissivity index ($\beta_{d}=2$ as in the case of the \citealt{Draine03} model) to directly
derive dust masses introduces an important bias, when $\beta$ is a variable parameter. However, we will show in the following that this is not really the case in this study.\\
We consider here two scenarios - one in which $\beta_{d}$ is a free parameter, while in the second one we fix the dust emissivity index to a value $\beta_{d}=2$, just as in other works.
This is done with the purpose of illustrating the differences that appear in the dust temperatures and the mean values of the sample, but also to investigate how fixing
$\beta_{d}$ can influence some of the relevant scaling relations involving dust temperature and SFR quantities.
Thus, to determine the best-fit values for ($T_{d}, \beta_{d}$) we start for each galaxy with two sets of test values for $k_{\nu}$ and $T_{d}$ in Eq.~\ref{eq:MBB}:
$k_{\nu}=[0.,...,10.] m^{2}/kg$ in steps of $0.05 m^{2}/kg$ and $T_{d}=[0.,...,80.] K$ in steps of 0.1K. In the first step, we determine the best-fit value for $T_{d}$ that
minimises the difference between the dust mass independently derived from Eq.~\ref{eq:Mdust_tauB} and the one determined using Eq.~\ref{eq:MBB} (the minimum of the $\chi^{2}$
function) considering the entire range of values for $k_{\nu}$ and $T_{d}$. With this best-fit value of the dust temperature - $T_{d}^{fit}$, in the second step, a best-fit 
blackbody function $B(\nu,T_{d}^{fit})$ is introduced now in Eq.~\ref{eq:MBB}, and a corresponding best-fit $k_{\nu}$ value is determined, $k_{\nu}^{fit}$. Finally, we derive
the dust emissivity index for each galaxy, $\beta_{d}$, using $k_{\nu}^{fit}$ and Eq.~\ref{eq:kappa_nu}.
In the second scenario, $\beta_{d}=2$ is fixed from the start, and using the same procedure we determine coresponding best-fit values for the dust temperature, $T_{d}^{bfix}$.
This procedure is similar with a grid method, to determine dust temperature values without considering any particular value for the dust emissivity index. This kind of procedure
should reduce spurious correlations between $T_{d}$ and $\beta_{d}.$ The consideration of a free $\beta_{d}$ was likewise used in other studies on NIR-FIR-submm data -
e.g. \cite{Boss12}, \cite{Gala12}, \cite{Smi12}, \cite{Cle13}, \cite{Cor14}, \cite{Remy13}, \cite{Taba14}, \cite{Gala14}, \cite{Kirk14}, \cite{Grossi15}, \cite{Lamp19}, \cite{Bendo24}. \\
Now, coming back to the issue of using a fixed $\kappa_{\nu_{0}}$ derived from a dust model calibrated on a fixed value of $\beta_{d}$, and following the procedure described
above, one can see that the potential bias introduced cancells out. First, this issue appeared in other studies when trying to derive \textit{temperature-dependent} dust masses
from SED fits (together with $T_{d}$ and $\beta_{d}$) while here we already have these masses derived - $M_{d}$. Second, $\kappa_{\nu_{0}}$ is present in both terms of the
minimising function, in both dust mass expressions from Eqs.~\ref{eq:Mdust_tauB} (embedded in the value of $K(B)$) \& \ref{eq:MBB}, and therefore when looking for the best-fit
value for $T_{d}$ that minimises the differences, its biasing effect is effectively cancelled. This is common for both analysed scenarios - variable and fixed $\beta_{d}$.

We also determine here the observed dust mass surface density, $\Sigma_{d}^{obs}$ essential in some of the investigated relations, using the observed dust disc scalength,
$R_{s}^{dust}$. As the diffuse dust is also considered to be distributed just as stars in an exponential disc following \cite{Pop11} model, the latter is determined from
the relation
\begin{eqnarray}\label{eq:dust_Rs}
R_{s}^{dust}=(M_{d}\kappa_{\nu}(B)/2\pi \tau_{B}^{f})^{1/2} 
\end{eqnarray}
with $\kappa_{\nu}(B)=2.46228 pc^{2}/M_{\odot}$ (dust mass absorbtion coefficient in B band) calculated from \cite{Draine03} model and their data tables, while for 
$\Sigma_{d}$ we have
\begin{eqnarray}\label{eq:sigma_dust}
\Sigma_{d}=M_{d}/2\pi (R_{s}^{dust})^{2}    
\end{eqnarray}
For two galaxies - NGC0024 and NGC4450 - dust temperatures and emissivity indices could not be derived as needed fluxes at 250$\mu m$ could not be identified in the
literature / databases, even though their dust opacities, dust masses and star-formation rates were calculated in Papers I and III. 

\subsection{Observed and corrected (intrinsic) dust properties}\label{sec:corr} 

Here we have to mention that in the following plots from Sect.~\ref{sec:results} and in Table~\ref{tab:temp_beta_Z_sigmadust}, one will see two sets of values for dust 
temperatures (in both scenarios - with $\beta_{d}$ variable or fixed), for $\beta_{d}$, and also for $R_{s}^{dust}$. For the dust temperatures and emisivity indices, this
may be counterintuitive. The explanation for this stands in the way these quantities are derived and the relations used in this process. Thus, dust temperatures are 
derived, as already metioned, based on dust masses independently derived through a self-consistent method in Paper I. One can see that the relations used for this purpose
- Eq.~\ref{eq:Grootes} \& \ref{eq:Mdust_tauB}, have quantities that depend on the observed B band disc scale-length, $R_{s,d}$, which is affected by dust, projection and 
decomposition effects (\citealt{Pas13a, Pas13b}). As a consequence, this resulted in two sets of values for dust masses, $M_{d}$ and $M_{d}^{i}$, the observed (measured) and
intrinsic (corrected) ones. Therefore, when applying the grid-like method described in Sect.~\ref{subsec:dust_temp_beta} and the minimisation function, one will obtain two
sets of values for the best-fit dust temperature, $T_{d}^{fit}$, one for each of $M_{d}$ and $M_{d}^{i}$. We call these observed and intrinsic dust temperatures, $T_{d}$ 
and $T_{d}^{i}$. This will create two different best-fit blackbody functions for each of these temperature values, and following the other steps of the procedure, two dust
emissivity index values - $\beta_{d}$ \& $\beta_{d}^{i}$. According with the naming pattern already used, we call these \textit{observed} and \textit{intrinsic} (corrected)
dust emissivity indices. For the same reasons, we have two sets of dust disc scalelengths as well, $R_{s}^{dust} \& R_{s}^{dust, i}$, as these are determined based on dust
masses - $M_{d}$ and $M_{d}^{i}$.\\
In the following section, we shall see from the plots and fits that there are some non-negligible differences for some relations and their characteristic parameters
when considering both sets of values for $T_{d}$ and $\beta$.

To calculate the intrinsic (corrected) dust surface densities, $\Sigma_{d}^{i}$ we use a similar formula as in Eq.~\ref{eq:sigma_dust} as follows
\begin{eqnarray}\label{eq:sigma_dust_corr}
\Sigma_{d}^{i}=M_{d}^{i}/2\pi (R_{s}^{dust, i})^{2}    
\end{eqnarray}
where $M_{d}^{i}$, the intrinsic dust masses of the galaxies, were previously derived in Paper I (see Table 2), and
\begin{eqnarray}\label{eq:dust_Rs_corr}
R_{s}^{dust, i}=(M_{d}^{i}\kappa_{\nu}(B)/2\pi \tau_{B}^{f})^{1/2}
\end{eqnarray}

\subsection{Error estimation}\label{sec:errors}

The uncertainties over $d_{gal}$ (measured distance to the galaxy) were taken from the same references as before (see Table 1 in Paper I). The uncertainties of the 250$\mu m$
fluxes and metallicities were taken from the references were the respecive values were found (Table~\ref{tab:temp_beta_Z_sigmadust}), while the ones for $M_{d}, \tau_{B}$
and $R_{s,d}(B)$ were already derived (also in Paper I).  We then performed propagation of errors in Eqs.~\ref{eq:MBB}-\ref{eq:dust_Rs_corr} to obtain the standard 
deviation ($\sigma$) for all the needed parameters ($T_{d}, \beta_{d}, R_{s}^{dust}, \Sigma_{d}$), both the observed and intrinsic quantities. The values are also shown in
Table~\ref{tab:temp_beta_Z_sigmadust}.

\section{Results}\label{sec:results}
 
\begin{figure}
 \begin{center}
 \includegraphics[scale=0.45]{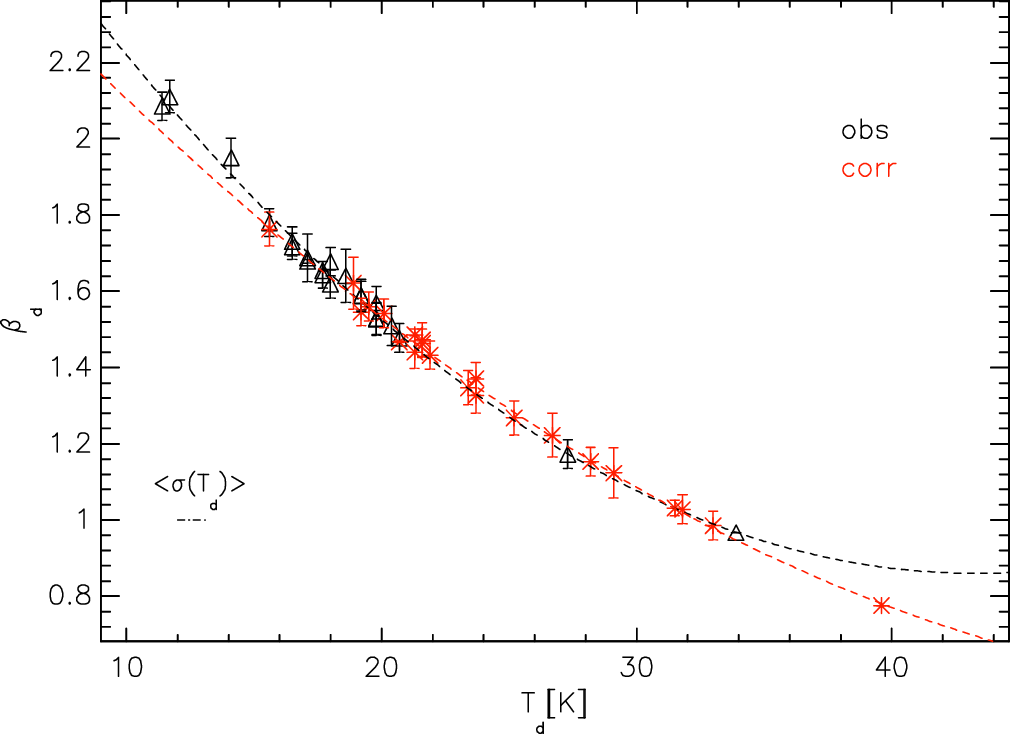}
 \caption{\label{fig:Tdust_beta_evol} The dust temperature ($T_{d}$) - dust emissivity index evolution ($\beta_{d}$). The observed $T_{d}$ are shown with black triangles,
 while the corrected ones are represented with red stars. The black (red) dotted curves are polynomial regression fits of second order to the observed (corrected) values.
 The error bars represent the standard deviations. The average standard deviation of $T_{d}$ is overplotted in the lower left corner.}
 \end{center}
\end{figure}
 
We show here the main results of this study - an analysis of the relations between the dust temperature, its emissivity index, $\beta$ (Sect.~\ref{subsec:Tdust_beta}), and local /
integrated ISM and SFR related quantities (Sect.~\ref{subsec:T_dust_sigma_SFR_dust}-\ref{subsec:sigma_dust_relations}).
It is important to mention that all the plots show results for the whole sample considered - spirals and lenticulars. However, whenever necessary, we do underline the
differences in the numerical results appearing when considering only the 19 spirals in our sample.\\ 
Throughout this section, some of the analysed relations are plotted in the linear form $log(Y) - log(X)$. The best-fit for each relation (from a linear
regression procedure) has the general form $log(Y)=\beta+\alpha \times log(X)$, with $\alpha$ - the intercept and $\beta$ - the slope of the relation. Unless specified
otherwise, all the intercepts and slopes are given in the same units as the ones of $log(Y)$ and $log(Y)/log(X)$.

\subsection{The dust temperature - $\beta_{d}$ evolution}\label{subsec:Tdust_beta}  

In Figure \ref{fig:Tdust_beta_evol} we show the plotted variation of dust temperature with dust emissivity index. It can be observed the known anti-correlation between
$T_{d}$ and $\beta_{d}$, with its hyperbolic shape seen in many other studies. It is not yet clear if it has, at least in part, a physical origin (\citealt{Shetty09b, 
Smi12, Gala12, Juv13, Remy13} or if it appears as a result of the $\chi^{2}$ fitting technique (\citealt{Shetty09a}). Other potential causes for this anti-correlation mentioned
in the literature are the noise in the NIR-FIR-submm data and temperature mixing along the line of sight (\citealt{Shetty09a, Shetty09b, JuvYsa12, Juv13, Taba14}), or the
use of a single-temperature MBB - that does not account for a warm dust component to characterise the cold dust emission SED (\citealt{Shetty09a, Kelly12, JuvYsa12, Juv13}),
and can produce biased values for $\beta_{d}$ (\citealt{Kirk14, Hunt15, Bendo23}). We have fitted the data points with a second-order polynomial to better describe the 
hyperbolic trend. On a log($T_{d}$)-log($\beta_{d}$) representation, the trend becomes linearly decreasing, and we get the best-fit relations for the observed and intrinsic 
values as
\begin{eqnarray}
log(\beta_{d})=(-0.70\pm0.03)log(T_{d})+(1.09\pm0.04)
\end{eqnarray}
\begin{eqnarray}
log(\beta_{d}^{i})=(-0.88\pm0.04)log(T_{d}^{i})+(1.32\pm0.03)
\end{eqnarray}
These relations are equivalent with $\beta_{d}=1.51(T_{d}/20)^{-0.70}$ and $\beta_{d}=1.55(T_{d}/20)^{-0.88}$, for an easier comparison with other results in the 
literature, that give this relation in the form $\beta_{d}=C(T_{d}/20)^{\alpha}$ (with C a constant). Compared with other studies, these values for the slope of the log-linear
relation characterise a shallower relation. For example, \cite{Para10} determined a relation $\beta_{d}\propto T_{d}^{-1.33}$ for their analysed Herschel PACS and SPIRE maps,
\cite{Smi12} derived slopes of $-1.57$ and $-0.61$ for the outer and inner regions of their sample, \cite{Grossi15} found a relation of $\beta_{d}\propto T_{d}^{-1.55}$ for
their sample of Virgo dwarf galaxies while \cite{Remy13} derived $T_{d}\propto \beta_{d}^{-0.29}$ for the KINGFISH sample of spirals and S0s. \\
We have to underline here that despite using dust masses not derived from a SED fit but through another independent method, and without having to deal with the noise in all
the observed NIR-FIR-submm fluxes (as unlike in other studies, we only use the 250$\mu m$ fluxes), we still observe the $T_{d}-\beta_{d}$ degeneracy. So, while we have eliminated
some of the causes for this degeneracy, it is still present, as we are still using a single-temperature MBB to characterise the dust emission overall in the galaxy. Likewise, 
by using an hierarchical Bayesian approach to fit the FIR SED, \cite{Lamp19} have only managed to reduce the degree of this anti-correlation and diminish the ranges of 
values for $T_{d}$ and $\beta_{d}$, but not entirely eliminate it. Therefore, we have reasons to believe that this anti-correlation may still, at least in part, have a 
physical origin, and this cannot be completely avoided with the current methods used in SED fitting and dust temperature / dust mass calculation. \\
For the purpose of comparison with other similar relevant studies, it is important to mention here the mean values calculated for the dust temperature and emissivity index
of the entire sample, in both cases. We actually derive these values for the spirals and S0s taken together (the whole sample), but also for the spiral galaxies only, for a
more relevant comparison with other results. Thus, in the case when $\beta_{d}$ was left free, we determined for the whole sample mean values of $<T_{d}>=17.08\pm0.33K$ and 
$<T_{d}^{i}>=22.40\pm0.81K$ for dust temperatures, and $<\beta_{d}>=1.63\pm0.05$ and $<\beta_{d}^{i}>=1.34\pm0.05$. The mean dust temperature values are very consistent with values
derived in \cite{Ski11, Gala12, Gro13, Remy13, Gala14, Grossi15, Hunt15}. If we consider only the 19 spirals in the sample, then the values change, not dramatically though:
$<T_{d}>=15.44\pm0.26K$ and $<T_{d}^{i}>=20.59\pm0.58K$, while for dust emissivity index we obtain $<\beta_{d}>=1.70\pm0.04$ and $<\beta_{d}^{i}>=1.39\pm0.04$. These values
are also in very good agreement with the previously mentioned studies, where the methods were sometimes different. For the second case, with \textit{fixed} $\beta_{d}$, the
mean dust temperatures for the whole sample are $<T_{d}^{bfix}>=20.12\pm 0.28K$, $<T_{d}^{i,bfix}>=30.83\pm 0.69K$, while for the spiral galaxies only, we have
$<T_{d}^{bfix}>=17.05\pm 0.23K$, $<T_{d}^{i,bfix}>=26.59\pm 0.50K$. We can see right away that when $\beta_{d}$ is constrained overall to a value higher than the one obtained
in the $\beta_{d}$-free scenario, both the observed and corrected average dust temperatures of the whole sample, but also of the spirals sub-sample, are \textit{systematically higher}. \\ 
All the numerical results for $T_{d}, \beta_{d}$ an their standard deviations ($\sigma$) can be found in Table~\ref{tab:temp_beta_Z_sigmadust}.

\begin{figure*}
\begin{center}
\hspace{-0.7cm}
\includegraphics[scale=0.26]{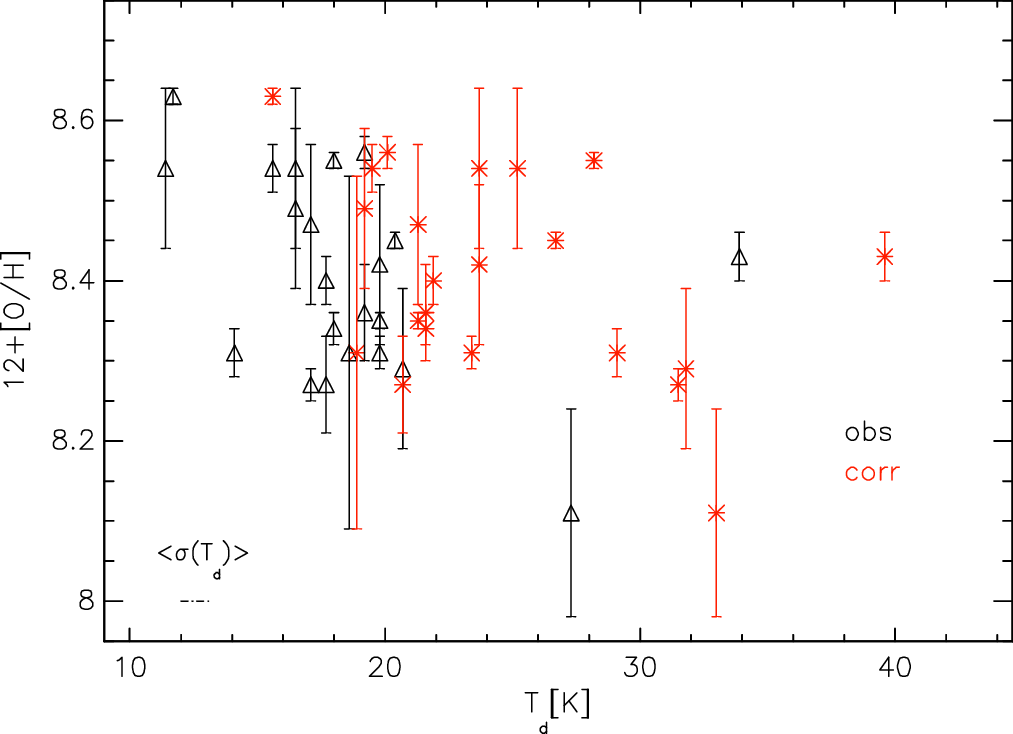}
\hspace{-0.08cm}
\includegraphics[scale=0.26]{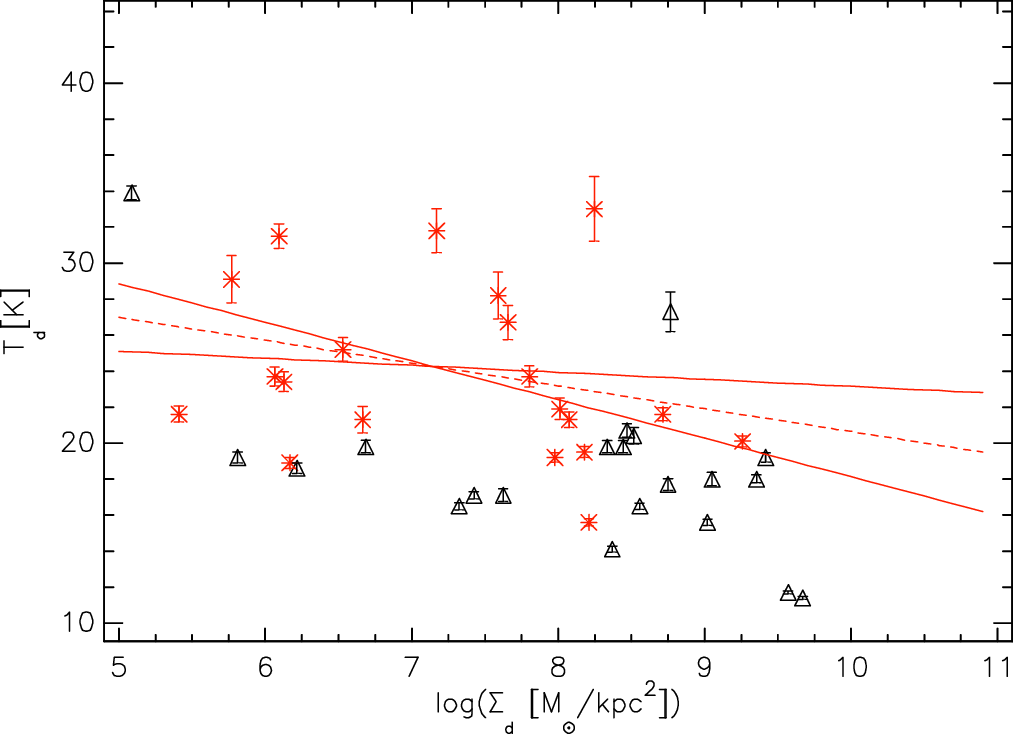}
\hspace{-0.08cm}
\includegraphics[scale=0.26]{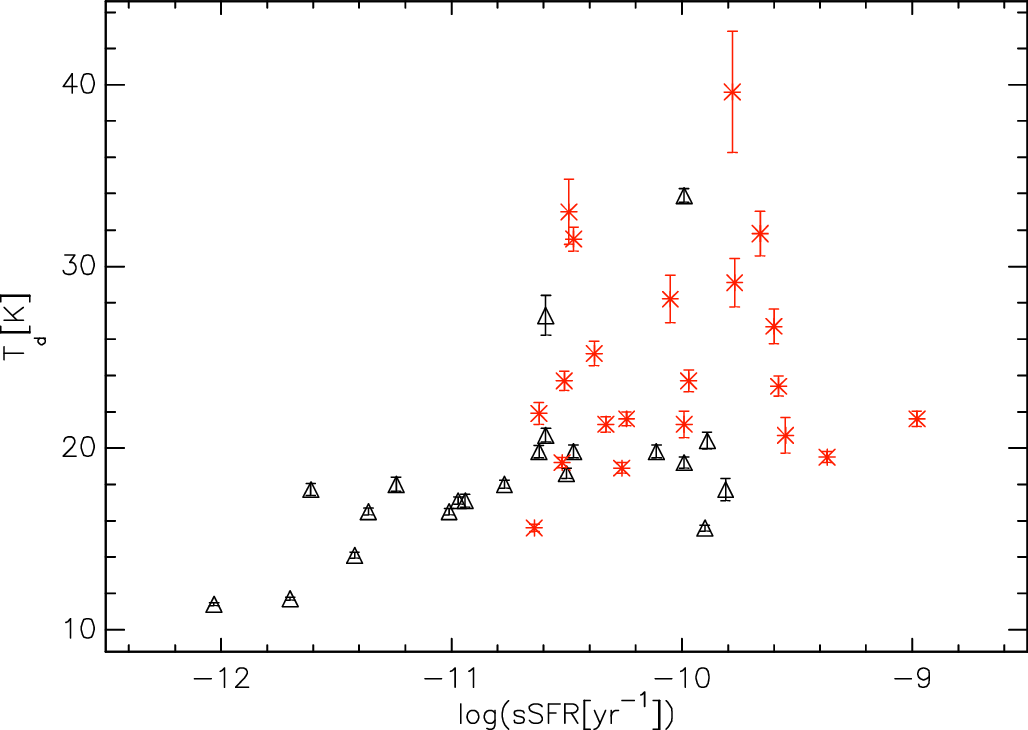}
\hspace{-0.08cm}
\includegraphics[scale=0.26]{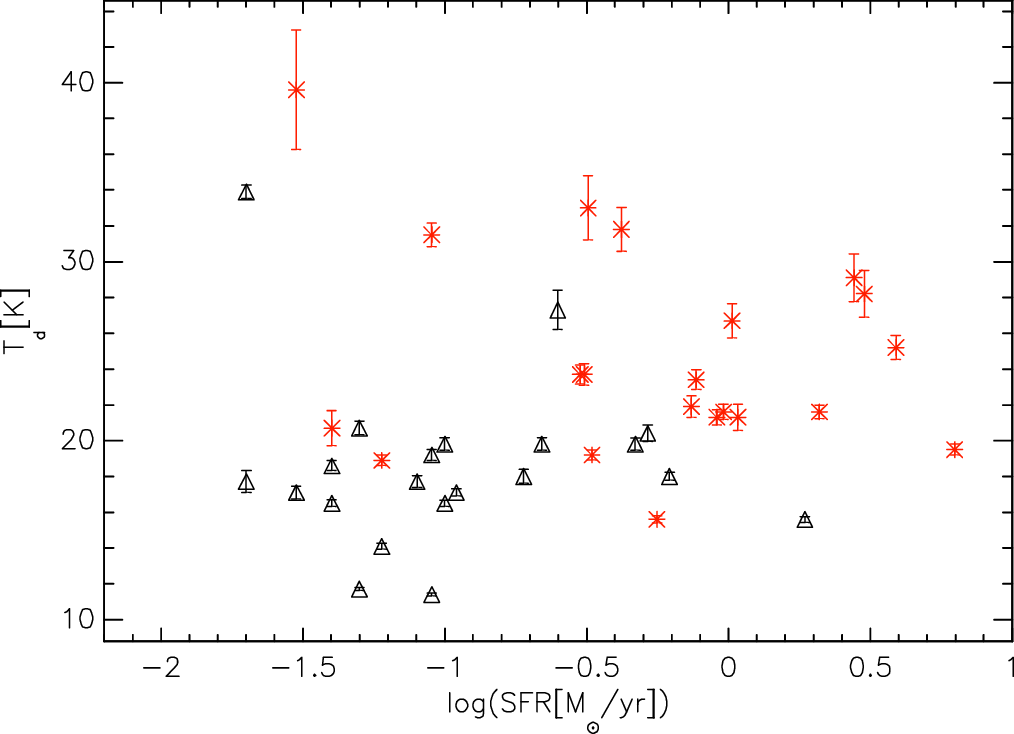}
\vspace{0.20cm}
\hspace{-0.7cm}
\includegraphics[scale=0.26]{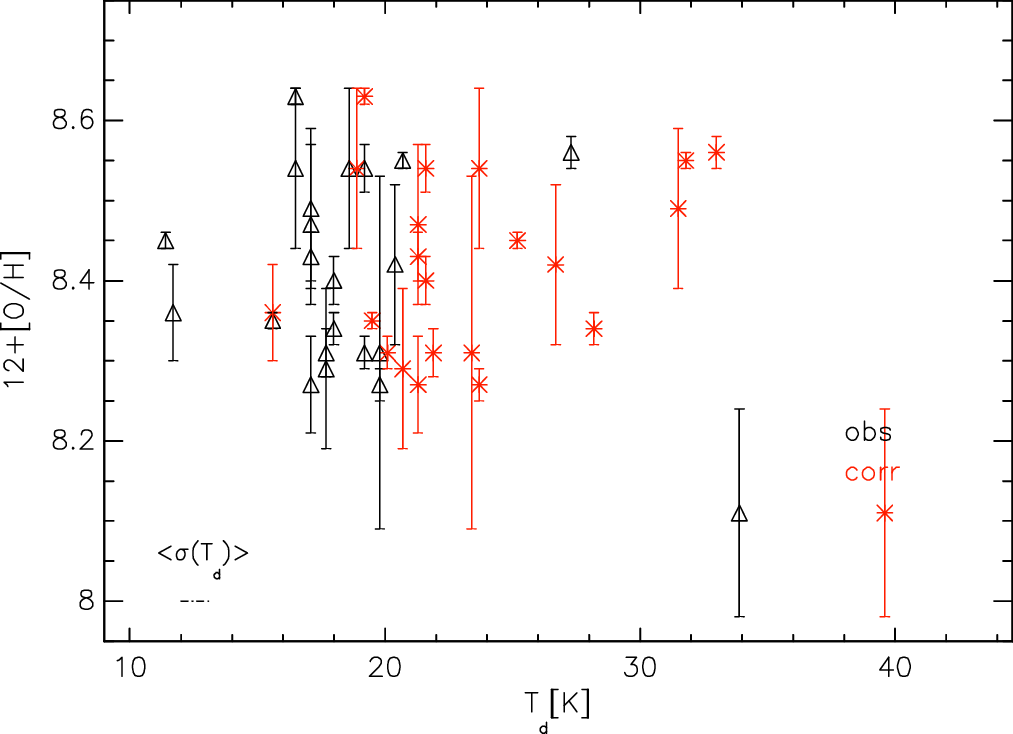}
\hspace{-0.08cm}
\includegraphics[scale=0.26]{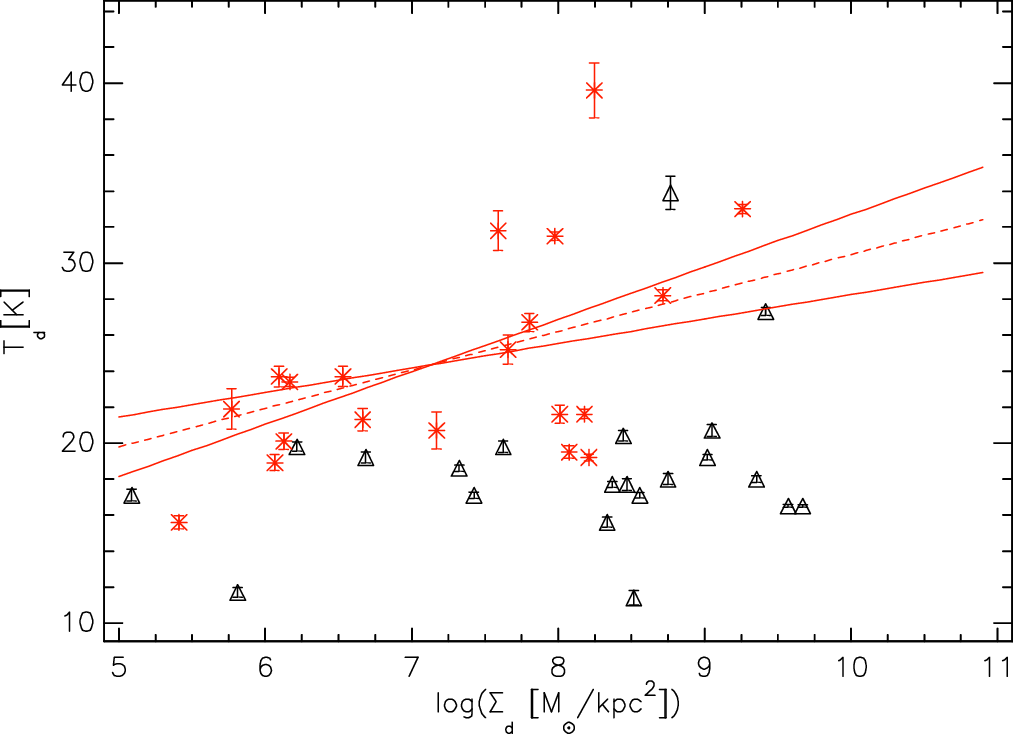}
\hspace{-0.08cm}
\includegraphics[scale=0.26]{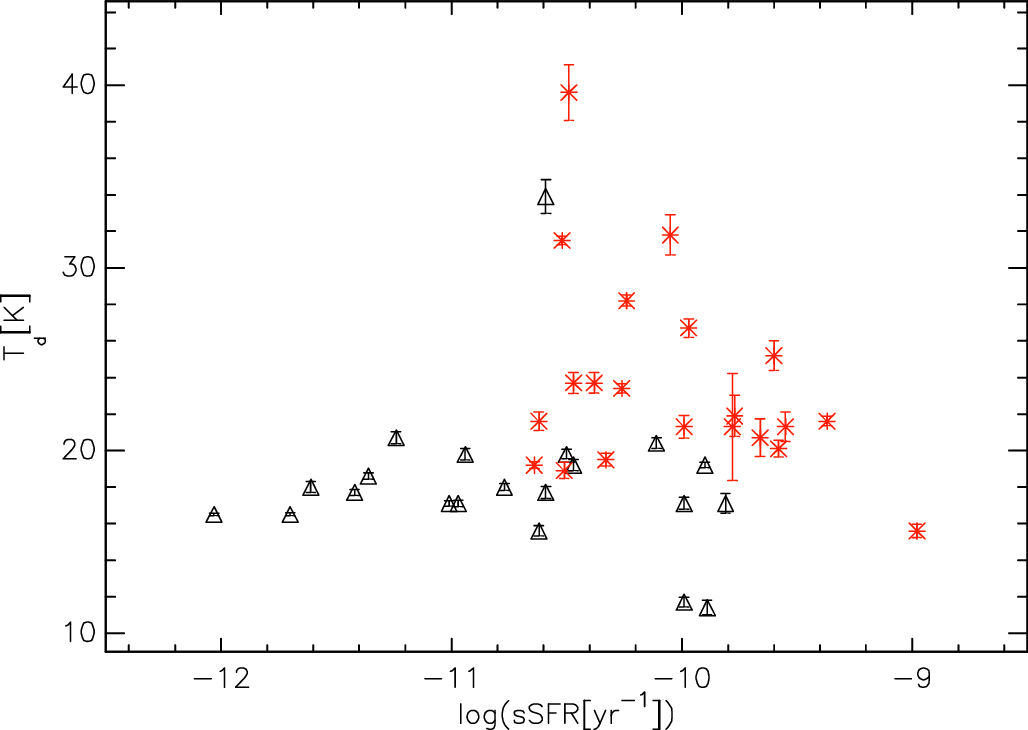}
\hspace{-0.08cm}
\includegraphics[scale=0.26]{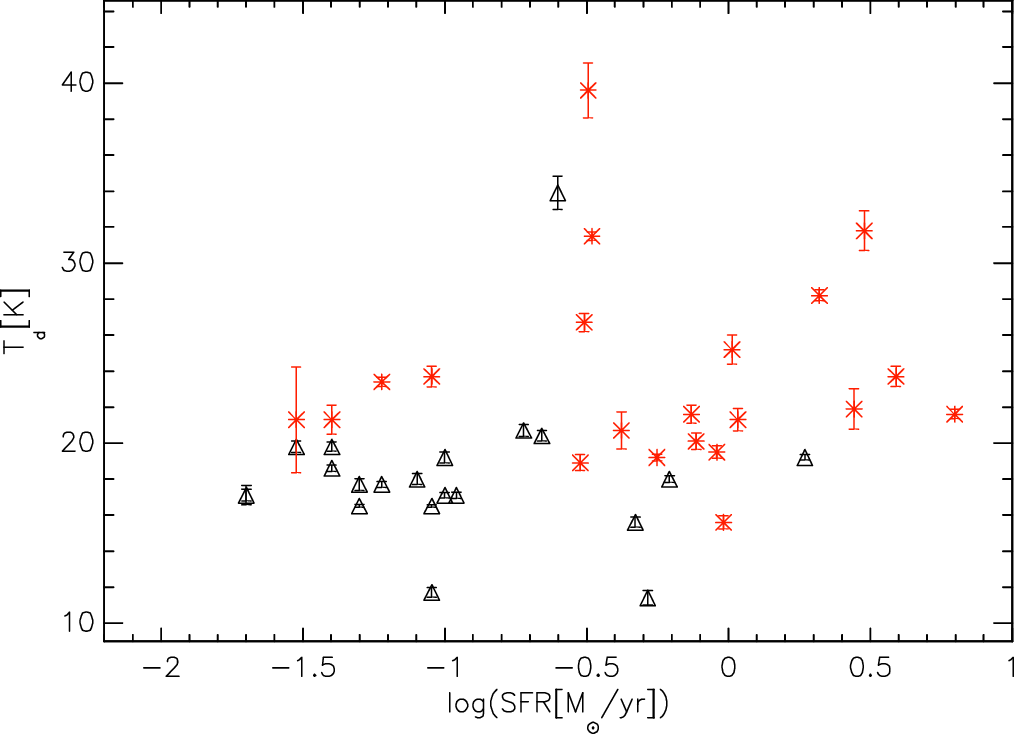}
\caption{\label{fig:Tdust_compare_plots_beta_free_vs_fixed} A comparison of the relations between the dust temperature ($T_{d}$) and some of the relevant dust/ISM and star-formation
related quantities, when $\beta_{d}$ is either free or fixed to a value of $2.0$. \textit{Upper row}: Dust temperature vs. metallicity (Z[O/H], taken from \citealt{Mous10}), 
dust surface density ($\Sigma_{d}$), sSFR and SFR, when $\beta_{d}$ is left free; \textit{Lower row}: the same plots for a
fixed $\beta_{d}=2$ value. The red dotted line is a linear regression fit of the corrected values, while the two red solid ones delimit the $\pm1\sigma$ uncertainty range
for the best-fit relation. The error bars represent the standard deviations. The average standard deviations of the abscisa quantities are overplotted on the figures.}
\end{center}
\end{figure*}

\subsection{$T_{d}$ - ISM/SF local/integrated quantities scaling relations}\label{subsec:T_dust_sigma_SFR_dust}

In the following figure - Fig.~\ref{fig:Tdust_compare_plots_beta_free_vs_fixed}, we plot the variation of dust temperature with metallicity (in the form of oxygen abundance,
12+[0/H]), with dust surface density ($\Sigma_{d}$), but also with some of the star-formation related quantities, namely SFR and specific star-formation, sSFR. The later plots
are produced to investigate if any of the SF quantities is a potential dust temperature tracer. As a means to compare the trends seen in the data and potential differences
that may arise, we show on the upper row of Fig.~\ref{fig:Tdust_compare_plots_beta_free_vs_fixed} the plots for the case where $\beta_{d}$ was left free, while the lower 
row of plots corresponds to the case with $\beta_{d}=2$. In the first pair of plots, we see the metallicity (taken from \citealt{Mous10}) variation with dust temperature.
A slightly decreasing trend can be noticed when $\beta_{d}$ is variable, for both the observed and corrected values, despite the large uncertainties for certain galaxies.
A part of the spread in the data can be due to the use of the oxygen abundance as a tracer of the total metal mass in these galaxies.
A similar result, though not very concludent, was observed by \cite{Remy13} in their analysis of KINGFISH galaxies, while \cite{Cor14} obtained a correlation coefficient 
of $-0.53$ for the same anti-correlation, in the case where they left $\beta_{d}$ as a free parameter.  For the second case, an inconclusive flat-like trend (as also noticed
by \citealt{Cor14} when fixing $\beta_{d}$ to 2) can be seen, different from the first scenario. A basically flat trend was also found by \cite{Lamp19} for both their 
approaches - a single MBB and a Bayesian hierarchical SED fitting. The decreasing trend suggests that metal-rich galaxies have colder dust on average. This translates into
important information about the dust grain composition and their properties.

On the second column in Fig.~\ref{fig:Tdust_compare_plots_beta_free_vs_fixed} we show the $T_{d}$ vs dust surface density variation. A relation between these quantities 
would be expected if the population of young stars is the main contributor to the dust heating. A mild anti-correlation is noticeable when $\beta_{d}$ is a free
parameter, if we disregard some obvious outliers, with a slope $\alpha=-1.27\pm0.87$ (corrected relation). A large scatter is present in both the observed and intrinsic values. This result means that colder dust is more concentrated in the disc of the galaxies
and therefore has a more compact distribution. However this comes in opposition with the positive strong correlation found by \cite{Tai25} from their pixel-by-pixel analysis
of a sample of spiral galaxies (a few also present in our sample), with correlation coefficients of 0.6 or higher. In the lower plot of the same column, we can see that
when $\beta_{d}=2$, the trend is reversed and we observe a mild increasing behaviour between $T_{d}$ and $\Sigma_{d}$, with a slope of $\alpha=1.29\pm0.67$ for the corrected values. \\
In the third and fourth columns of the same figure, we show $T_{d}$ variation with specific star-formation and the star-formation rate, SFR, for both cases. The trends in 
the plots are rather inconclusive, with a large scatter in the data. Fixing $\beta_{d}$ does not produce more clear results. The same situation is for the $T_{d}-\Sigma_{SFR}$ variation, which we do not show here. There is a
weak increasing tendency between $T_{d}$ and sSFR, for the $\beta_{d}$-\textit{free} case. A more clear increasing trend was observed for the relation between $T_{d}$, sSFR
and $\Sigma_{SFR}$ by \cite{Chiang23}, while \cite{Lamp19} found correlation coefficients of 0.54 and 0.49 for the $T_{d}-sSFR$ and $T_{d}-\Sigma_{SFR}$ relations. Our results
suggest that sSFR, SFR, $\Sigma_{SFR}$ are not conclusive tracers for the dust temperature. This result can be caused by either the low statistics of this study or the 
consideration of a single temperature MBB, which does not take into account a warmer dust component, present in the star-forming regions around the young stars.\\
While we do not show it here, we have to mention also that no increase or decrease of the dust-to-stellar mass ratio ($M_{d}/M_{\star}$) with $T_{d}$ could be noticed, the
trend being basically flat for both cases, with some scatter in the data.

\begin{figure*}
\begin{center}
\hspace{-0.7cm}
\includegraphics[scale=0.26]{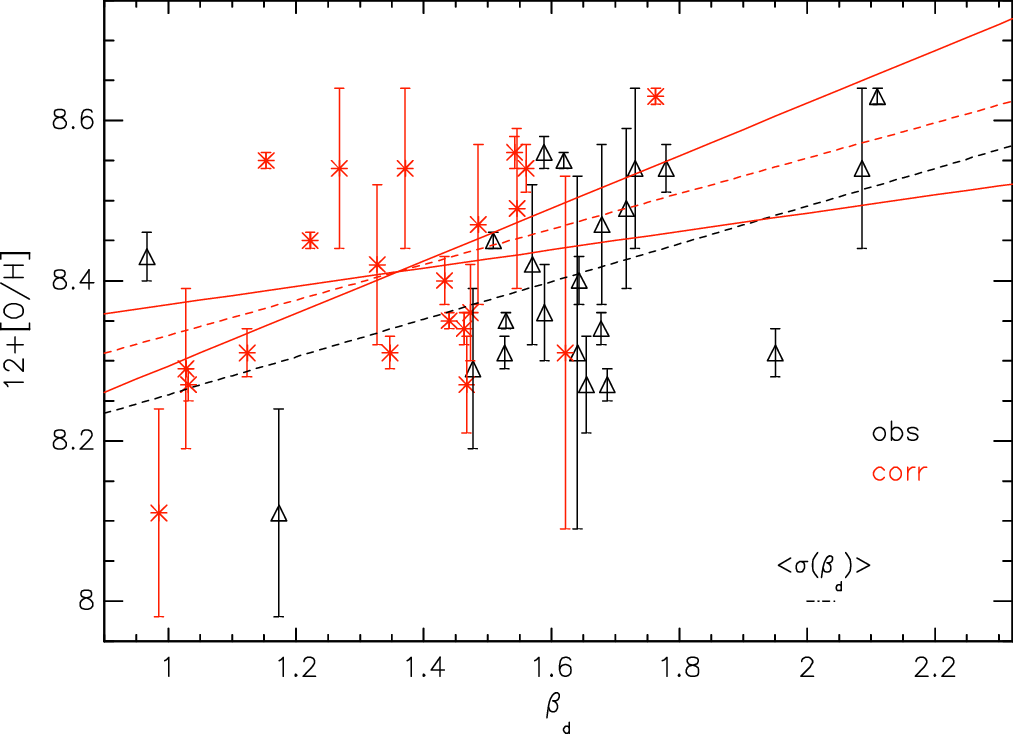}
\hspace{-0.08cm}
\includegraphics[scale=0.26]{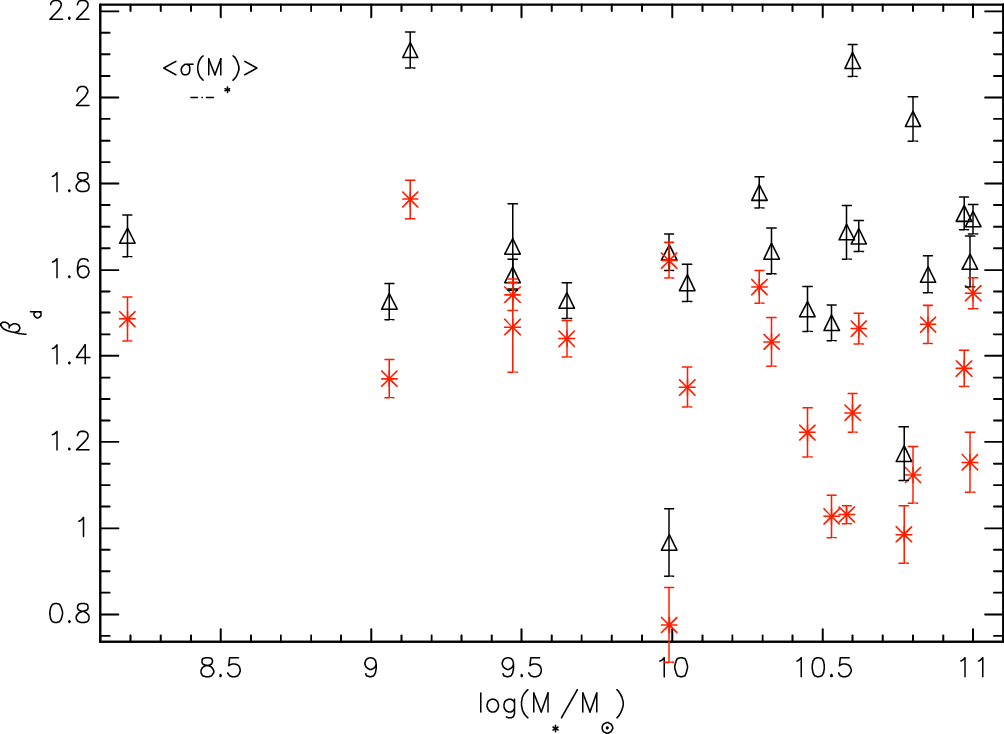}
\hspace{-0.07cm}
\includegraphics[scale=0.26]{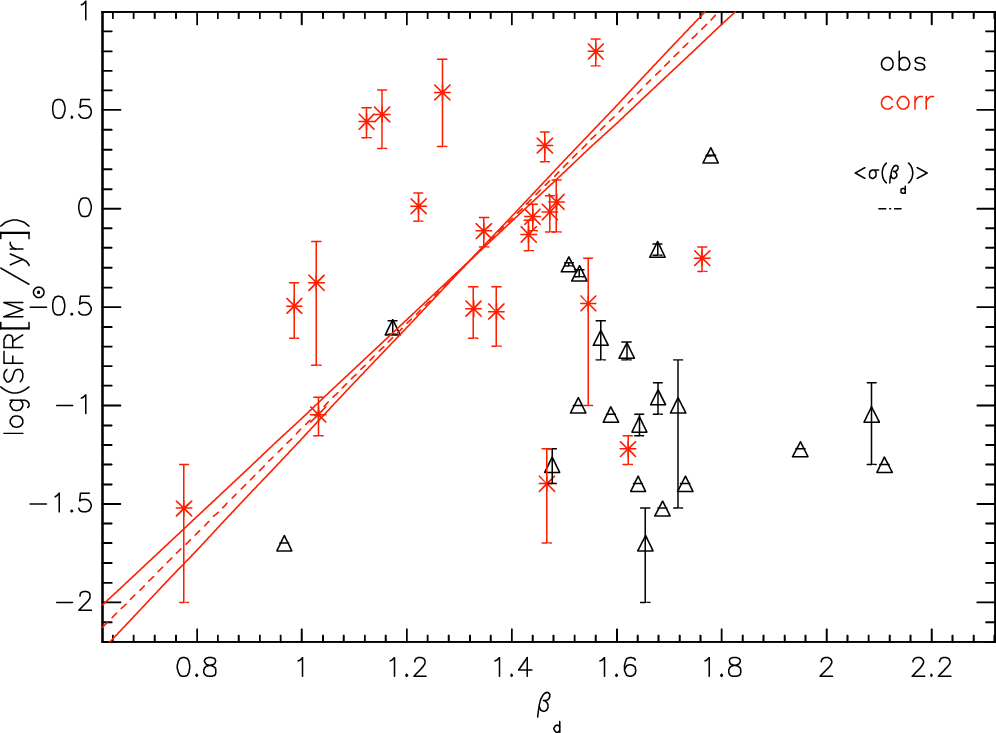}
\hspace{-0.07cm}
\includegraphics[scale=0.26]{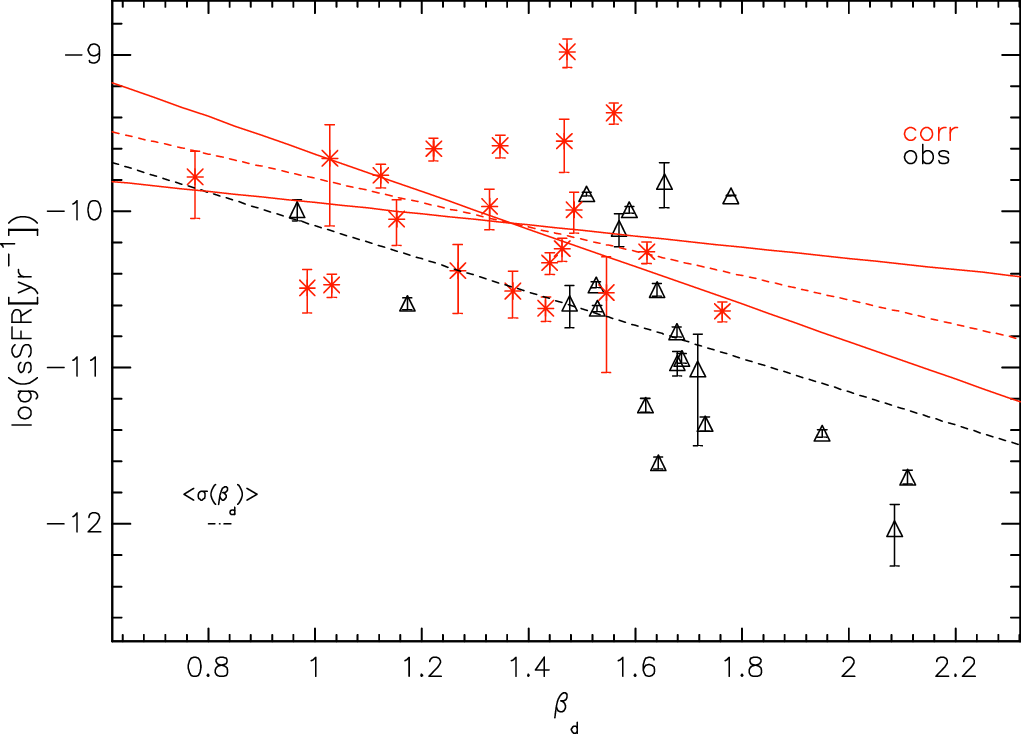}
\caption{\label{fig:beta_ZOH_ISM_quantities} The relation between the dust emissivity index, $\beta_{d}$, and other ISM/SFR quantities: metallicity (12+[O/H]), stellar mass
($M_{\odot}$), SFR and sSFR. The black dotted line is a linear regression fit of the observed values. The two red solid lines delimit the $\pm1\sigma$ uncertainty range
for the best-fit relation of the corrected values. The rest of symbols, colors and lines have the same meaning as those in Fig.~\ref{fig:Tdust_compare_plots_beta_free_vs_fixed}.}
\end{center}
\end{figure*}

\subsection{$\beta_{d}$ - ISM/SF local/integrated quantities scaling relations}\label{subsec:beta_dust_relations}

In this section we explored some of the potential relations of $\beta_{d}$ with relevant ISM and SF measurements. We show in Fig.~\ref{fig:beta_ZOH_ISM_quantities} some of
these relations. In the first panel, the variation of dust emisivity index is plotted against metallicity. A mild correlation can be observed, with slopes of $0.23\pm0.10$
and $0.22\pm0.11$ for the observed (corrected) relations, reduced level of scatter ($\sigma=0.21$ \& $0.17$ dex), while the derived corresponding Pearson correlation coefficients,
$r_{Z,\beta_{d}}$, have the values of 0.46 and 0.42. This relation has also been found in studies by \cite{Boss12, Remy13, Cor14, Lamp19}, with correlation coefficients around 
0.6 or higher (0.68 in \citealt{Cor14}). While the slope is quite shallow, this relation shows that metal-rich galaxies tend to have dust with a grain composition made 
predominantly from crystalline silicates, with less amorphous silicates present, as $\beta_{d}$ is a parameter known to be in connection with grain physical properties.
We should mention here that metallicity correlates even better with another property of dust, namely the dust mass. This relation, while not shown here, is a consequence
of the fundamental mass-metallicity and $M_{d}-M_{\ast}$ relations, and has a larger scatter and higher correlation coefficients that the $Z-\beta_{d}$, with $\sigma=0.70$,
$r_{M_{d},Z}=0.50$ for the intrinsic relation.\\
In the second panel we show how $\beta_{d}$ changes with the total stellar mass of the galaxy, $M_{\star}$. Whereas in some works (e.g. \citealt{Cor14, Lamp19}) a correlation
betwee these two quantities - a local and an integrated one - has been observed (although with a rather shallow slope), here for our small sample the results are inconclusive.
Thus, \cite{Lamp19} found a mid-strength correlation by analysing a sample of 192 galaxies from the JINGLE survey (\citealt{Saint18}), but also found a relation for $\beta_{d}-\mu_{*}$ and
$\beta_{d}-M_{HI}/M_{\odot}$ pairs ($M_{HI}$ is the mass of neutral atomic hydrogen), while \cite{Cor14} found a weak ($r=0.49$) correlation for $\beta_{d}-M_{\star}$, and
stronger ones for the other two (with coefficients r=0.65 and -0.70 respectively), when $\beta_{d}$ was left as a free parameter. We do not show here the other two plots
as again, a clear conclusive trend was not identified and the data were quite scattered.

The third and fourth panels show the variation of dust emisivity index with the star-formation rate and specific SFR. Here a significant scatter in the data stands out in
both plots, but nevetheless an increasing trend of SFR with $\beta_{d}$ for the corrected values only could be noticed, while an anti-correlation for both the observed and 
corrected values can be seen in the last plot. For the last relation we obtained the best-fit parameters: slope $\alpha=-1.10\pm0.66$, $r_{\beta_{d},sSFR}=-0.20$ - observed,
and $\alpha=-0.78\pm0.42$, $r_{\beta_{d},sSFR}=-0.76$, $\sigma=0.30$dex - for the corrected data. We see that galaxies with higher star-formation rates, and therefore more
active and with a higher dust production and dust mass ($M_{d}$ correlates with SFR, \citealt{Grossi15, Pas24}) have also higher dust emissivity indices, and therefore more
silicate rich grains. The slope of $2.66\pm0.16$ and the increasing trend for $SFR-\beta_{d}$ are not in agreement with the results of \cite{Lamp19} who did not observe a
conclusive increase between these quantities or between $\beta_{d}$ and sSFR. \cite{Smi12} investigated the $SFR-\beta_{d}$ relation but a clear correlation was not seen.
In our case, given the large scatter in both the observed and corrected data points for the last two plots, for a more clear and definitive conclusion on the $SFR-\beta_{d}$
and $sSFR-\beta_{d}$, a larger sample would be beneficial.

\begin{figure*}
\begin{center}
\hspace{-0.7cm}
\includegraphics[scale=0.26]{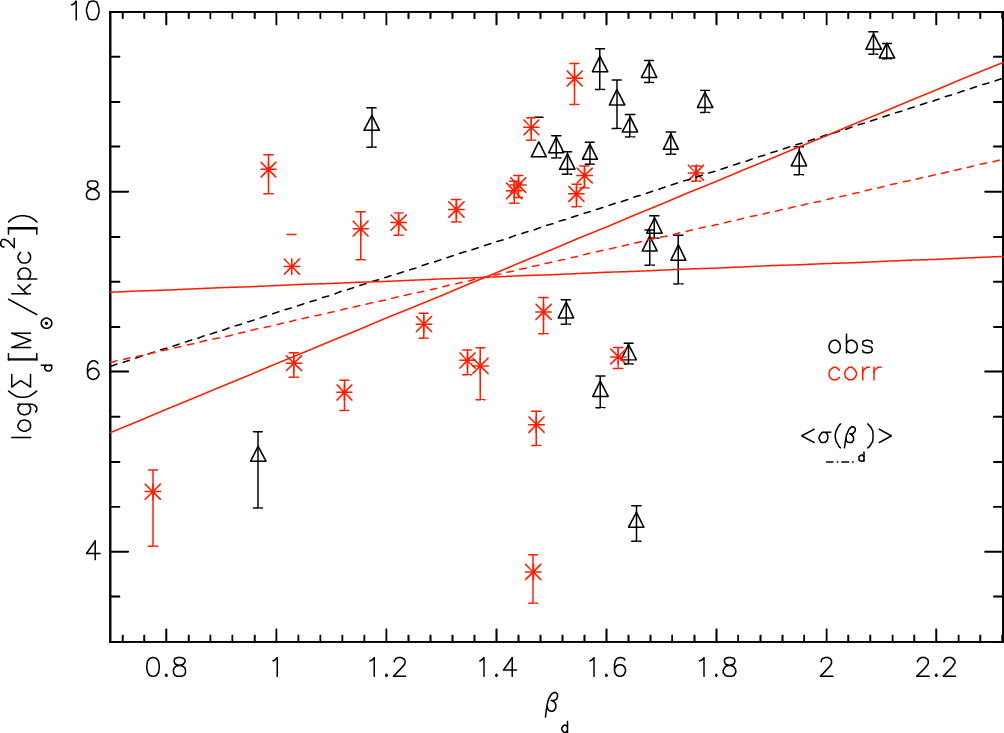}
\hspace{-0.08cm}
\includegraphics[scale=0.26]{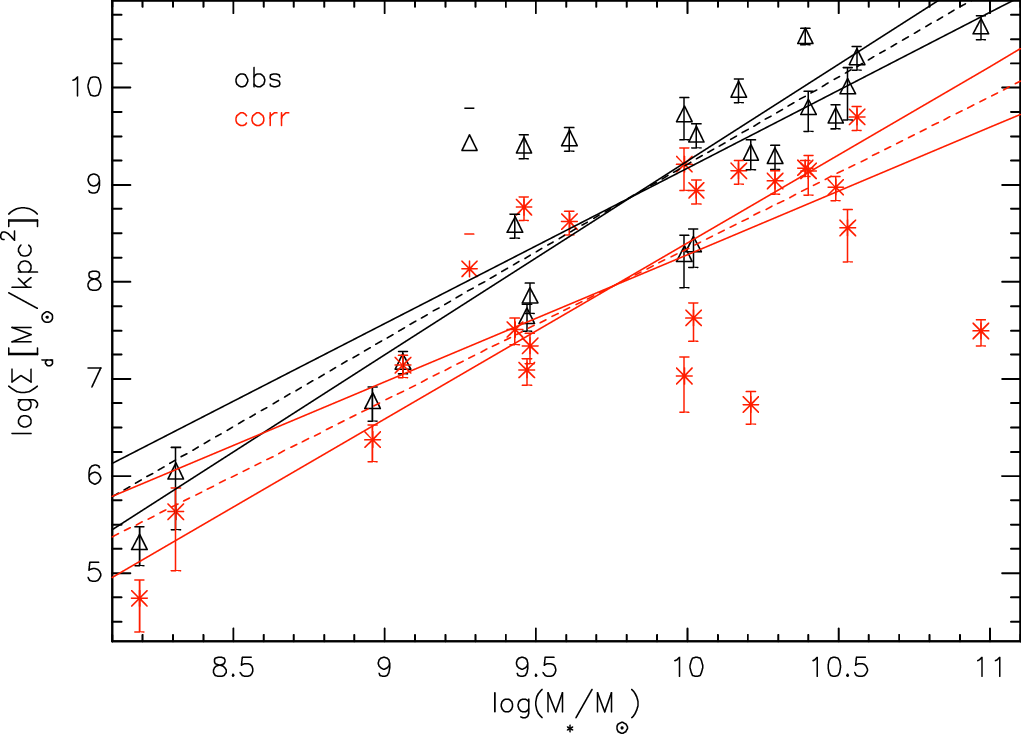}
\hspace{-0.07cm}
\includegraphics[scale=0.26]{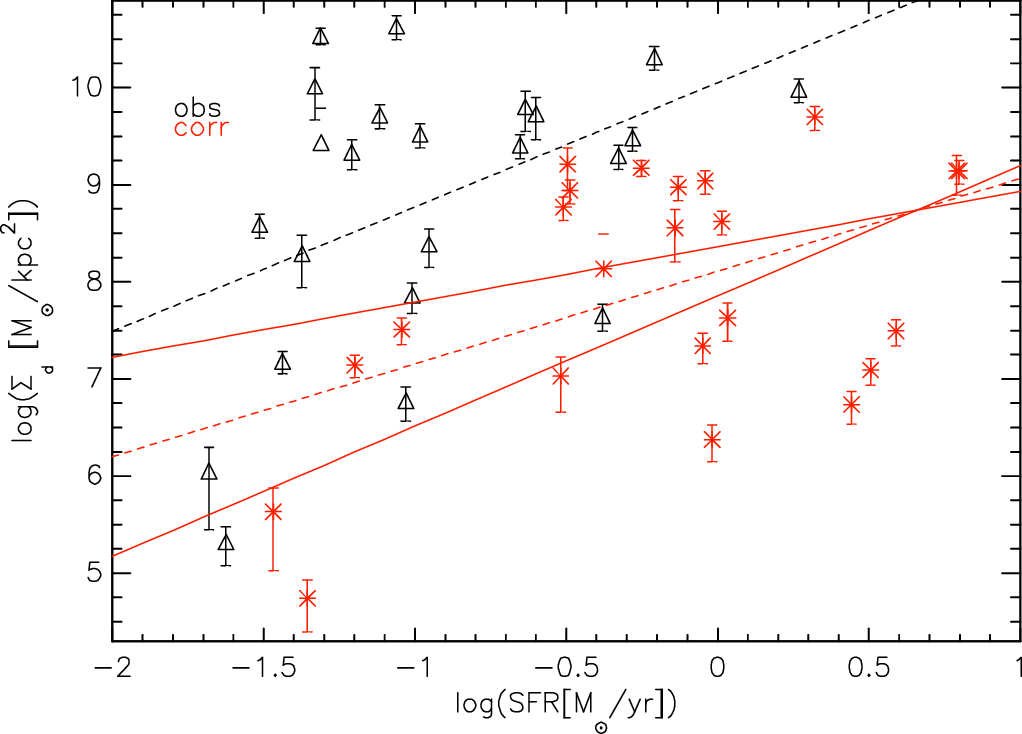}
\hspace{-0.07cm}
\includegraphics[scale=0.26]{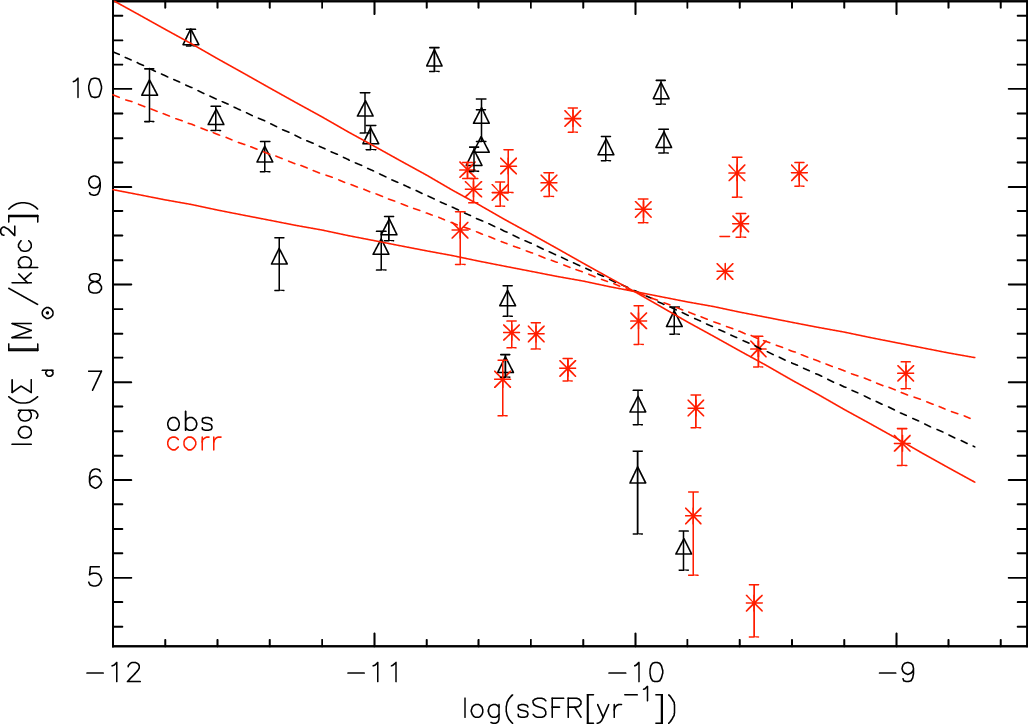}
\caption{\label{fig:sigma_dust_SFR_ISM_quantities} The relation between the dust surface density, $\Sigma_{d}$ and other dust/ISM/SFR quantities: $\beta_{d}$, stellar mass
($M_{\odot}$), SFR and sSFR. The black dotted line is a linear regression fit of the observed values, while the two black solid lines delimit the $\pm1\sigma$ uncertainty
range for the best-fit relation of the observed values. The rest of symbols, colors and lines have the same meaning as those in Fig.~\ref{fig:Tdust_compare_plots_beta_free_vs_fixed}.}
\end{center}
\end{figure*}

\subsection{$\Sigma_{d}$ - ISM/SF local/integrated quantities scaling relations}\label{subsec:sigma_dust_relations}

In this section, we investigated the relations between the dust surface density and some of the relevant dust, ISM and SF parameters - namely $\beta_{d}, M_{\star}, SFR$, and
$sSFR$. From the following Fig.~\ref{fig:sigma_dust_SFR_ISM_quantities}, one can see that the strongest correlation and lowest scatter ($\sigma$) for the dust surface density
is with the stellar mass (second panel). For this relation we derived correlation coefficients $r_{\Sigma_{d},M_{\star}}$ of 0.89 and 0.81 for the observed and corrected
values, which underlines the strength of this correlation. The corresponding slope and scatter for the corrected data points are $\alpha=1.56\pm0.25$ and $\sigma=0.63$dex,
respectively, with a slightly higher slope and comparable $\sigma$ for the measured values.\\
As $\Sigma_{d}\propto M_{d}$ (Eq.~\ref{eq:sigma_dust}), this relation can be considered a consequence / result of the more tight $M_{d}-M_{\star}$ relation (\citealt{Gro13, DeV17, vanG22, Pas24}).
To futher investigate this issue, we did a linear partial correlation analysis for the ($\Sigma_{d}, M_{\ast}, M_{d}$) quantities, deriving the corresponding partial
correlation coefficients: $r_{M_{\ast}\Sigma_{d}, M_{d}}=-0.105$, $r_{\Sigma_{d}M_{d},M_{\ast}}=0.673$ and $r_{M_{\ast}M_{d}, \Sigma_{d}}=0.747$. These values suggest that
$M_{d}$ is a more important quantity than $\Sigma_{d}$ for the stellar mass $M_{\ast}$. This was somehow expected as the $M_{d}-M_{\ast}$ correlation was found to be very 
strong and tight ($r_{M_{d},M_{\ast}}=0.92$ for the corrected values, low scatter $\sigma$). However, as our sample is small, better statistics would certainly strengthen
this conclusion.\\
In the first panel of the same figure we verified if there exists a relation between $\Sigma_{d}$ and $\beta_{d}$. We assumed there should be, as both of these quantities
are dust-related, even though one characterises an intrinsic property of the dust grains ($\beta_{d}$) while the other is a measure of the distribution of dust within the
associated disc ($\Sigma_{d}$). This could be investigated here as we considered $\beta_{d}$ as a free parameter, but not in other studies where it was fixed. One can see the large scatter
in both the observed and corrected data values. The derived correlation coefficients indicate a rather moderate correlation for the spirals, as $r_{\Sigma_{d}, \beta_{d}}=0.43$ for the 
observed values, and $0.42$ for the intrinsic ones. This relation is, at least according to our knowledge, \textit{new}, as it was not been observed in other studies.  It shows that dust grains with a higher composition of silicates can be found in galaxies with a more compact dust distribution, where processes of dust growth (through coagulation and material accretion) may be more prevalent, given the incresed shielding from the incoming radiation . \\
In the third panel, we plotted $\Sigma_{d}$ as a function of the star-formation rate. A relation between these two parameters would be expected given the $\tau-SFR$ correlation
found in Paper III (see there Fig. 6), the proportionality relation between $\tau$ and $\Sigma_{d}$ ($\tau=\kappa_{\nu}\Sigma_{d}$) and also because new stars are born in 
birthclouds of dust and gas. There is a large scatter in the data points (as in the case of the $\tau-SFR$ relation) and a mild increase in the dust surface density with SFR, with a rather shallow slope of $0.95\pm0.38$ for the corrected values. 
The derived Pearson coefficients are quite low - 0.45 \& 0.47, which also shows the mild strength of this relation. A strong correlation was also observed in \cite{Smi12}, for
the outer regions of their analysed galaxies. In comparison, the trend of $\Sigma_{d}$ with the specific
star-formation rate is considerably stronger, as it can be noticed in the last panel of Fig.~\ref{fig:sigma_dust_SFR_ISM_quantities}. The characteristic parameters of this
relation are: $\alpha=-1.01\pm0.48$ and $-1.22\pm0.36$ for the slopes (corrected / observed), $r_{\Sigma_{d},sSFR}=-0.42$ and $-0.59$ for the corresponding correlation coefficients.
The spread of the data points is still quite significant for this relation too. This last relation shows that dust is more compactly distributed in galaxies with lower sSFR, and 
therefore higher stellar mass and higher dust attenuation (given the known anti-correlation $sSFR-M_{\star}$ and the $M_{d}-M_{\star}$ correlation). \\
While we do not show it here, we note that $\Sigma_{d}$ and the stellar mass surface density ($\mu_{\star}$) are not correlated.

\begin{table*}
\caption{\label{tab:temp_beta_Z_sigmadust} Dust temperature and dust emissivity index for the whole sample. The different columns represent: (1) - galaxy name; (2) -
integrated flux at 250$\mu m$ taken from: $a$ - \protect\cite{Ani20}, $b$ - \protect\cite{Dale17}; $c$ - \protect\cite{OHal10}; (3) - dust temperature (free $\beta_{d}$);
(4) - intrinsic (corrected) dust temperature (free $\beta_{d}$); (5) - dust emissivity index; (6) - intrinsic (corrected) dust emissivity index; (7) -  dust temperature
(fixed $\beta_{d}$); (8) - corrected dust temperature (fixed $\beta_{d}$); (9) - observed dust surface density; (10) - intrinsic (corrected) dust surface density; (11) -
gas-phase metallicity (oxygen abundances), taken from: \protect\cite{Mous10}. In square brackets we have the units in which these quantities are given.}
\begin{tabular}{{r|r|r|r|r|r|r|r|r|r|r}}
\hline
 $Galaxy$ &  $f_{250}$ & $T_{d}$  &  $T_{d}^{i}$ & $\beta_{d}$  & $\beta_{d}^{i}$  &  $T_{d}$ & $T_{d}^{i}$ & $\Sigma_{d}$ &  $\Sigma_{d}^{i}$ & $12+[O/H]$ \\
          &     [Jy]   &   [K]    &    [K]         &           &         &   [K]    &     [K]        & $[\frac{M_{\odot}}{kpc^{2}}]$ & $[\frac{M_{\odot}}{kpc^{2}}]$  &     \\
     (1)  & (2)  &  (3)  &  (4) &  (5)  &  (6)  &  (7)  &  (8)  &  (9)  & (10)  & (11)\\
\hline
NGC 0628  &       $61.00\pm4.70^{a}$   & 19.80$\pm$0.34 &  21.30$\pm$0.42 &   1.53$\pm$0.04 &   1.44$\pm$0.04   &   5.60$\pm$0.29 &  19.50$\pm$0.29  &  8.33$\pm$7.78 &    8.07$\pm$7.52  &  8.35$\pm$0.01  \\
NGC 2841  &       $33.90\pm2.50^{a}$   & 15.60$\pm$0.16 &  19.50$\pm$0.30 &   1.78$\pm$0.04 &   1.56$\pm$0.04 	&  19.20$\pm$0.15 &  21.60$\pm$0.15  &  9.02$\pm$8.46 &    8.18$\pm$7.62  &  8.54$\pm$0.03  \\
NGC 2976  &       $24.60\pm1.80^{a}$   & 19.20$\pm$0.32 &  21.60$\pm$0.44 &   1.59$\pm$0.04 &   1.47$\pm$0.04   &  11.70$\pm$0.27 &  15.60$\pm$0.27  &  5.81$\pm$5.39 &    5.41$\pm$5.02  &  8.36$\pm$0.06  \\
NGC 3031  &      $176.00\pm13.00^{b}$  & 11.70$\pm$0.08 &  15.60$\pm$0.20 &   2.11$\pm$0.04 &   1.76$\pm$0.04   &  16.50$\pm$0.09 &  19.20$\pm$0.09  &  9.57$\pm$8.86 &    8.21$\pm$7.50  &  8.63$\pm$0.01  \\
NGC 3190  &        $8.50\pm0.58^{a}$   & 16.50$\pm$0.18 &  19.20$\pm$0.27 &   1.72$\pm$0.03 &   1.55$\pm$0.04   &  17.10$\pm$0.16 &  31.50$\pm$0.16  &  8.56$\pm$8.00 &    7.98$\pm$7.42  &  8.49$\pm$0.10  \\
NGC 3621  &       $37.00\pm4.60^{a}$   & 17.10$\pm$0.35 &  31.50$\pm$0.67 &   1.69$\pm$0.06 &   1.03$\pm$0.02 	&  19.80$\pm$0.31 &  23.70$\pm$0.31  &  7.62$\pm$7.07 &    6.09$\pm$5.58  &  8.27$\pm$0.02  \\
NGC 3938  &       $22.90\pm1.90^{a}$   & 19.80$\pm$0.34 &  23.70$\pm$0.60 &   1.57$\pm$0.04 &   1.33$\pm$0.05 	&  20.40$\pm$0.30 &  26.70$\pm$0.30  &  8.44$\pm$7.89 &    7.80$\pm$7.25  &  8.42$\pm$0.10  \\
NGC 4254  &       $62.00\pm6.00^{a}$   & 20.40$\pm$0.46 &  26.70$\pm$0.95 &   1.51$\pm$0.05 &   1.22$\pm$0.06 	&  11.40$\pm$0.39 &  25.20$\pm$0.39  &  8.52$\pm$7.96 &    7.66$\pm$7.10  &  8.45$\pm$0.01  \\
NGC 4594  &       $23.50\pm1.60^{a}$   & 11.40$\pm$0.07 &  25.20$\pm$0.66 &   2.09$\pm$0.04 &   1.27$\pm$0.05 	&  16.50$\pm$0.08 &  23.70$\pm$0.08  &  9.67$\pm$9.11 &    6.53$\pm$6.02  &  8.54$\pm$0.10  \\
NGC 4736  &       $65.00\pm7.00^{a}$   & 14.10$\pm$0.17 &  29.10$\pm$1.33 &   1.95$\pm$0.05 &   1.12$\pm$0.07 	&  17.70$\pm$0.16 &  21.90$\pm$0.16  &  8.37$\pm$7.90 &    5.77$\pm$5.34  &  8.31$\pm$0.03  \\
NGC 4826  &       $42.40\pm3.00^{a}$   & 16.50$\pm$0.19 &  23.70$\pm$0.53 &   1.73$\pm$0.04 &   1.37$\pm$0.04 	&  18.60$\pm$0.17 &  18.90$\pm$0.17  &  7.32$\pm$7.07 &    6.06$\pm$5.83  &  8.54$\pm$0.10  \\
NGC 5055  &      $138.00\pm9.00^{a}$   & 17.70$\pm$0.33 &  21.90$\pm$0.59 &   1.64$\pm$0.05 &   1.43$\pm$0.06 	&  18.00$\pm$0.29 &  21.60$\pm$0.29  &  8.75$\pm$8.19 &    8.01$\pm$7.45  &  8.40$\pm$0.03 \\
NGC 5474  &        $9.13\pm0.43^{a}$   & 18.60$\pm$0.29 &  18.90$\pm$0.30 &   1.64$\pm$0.04 &   1.62$\pm$0.04 	&  19.80$\pm$0.25 &  23.40$\pm$0.25  &  6.22$\pm$5.64 &    6.17$\pm$5.59  &  8.31$\pm$0.22 \\
NGC 7331  &       $88.20\pm6.20^{a}$   & 18.00$\pm$0.22 &  21.60$\pm$0.36 &   1.68$\pm$0.04 &   1.46$\pm$0.04 	&  18.00$\pm$0.20 &  28.20$\pm$0.20  &  9.35$\pm$8.79 &    8.71$\pm$8.15  &  8.34$\pm$0.02 \\
NGC 7793  &       $54.00\pm3.70^{a}$   & 19.80$\pm$0.35 &  23.40$\pm$0.55 &   1.53$\pm$0.04 &   1.35$\pm$0.04 	&  19.20$\pm$0.30 &  20.10$\pm$0.30  &  6.69$\pm$6.17 &    6.13$\pm$5.61  &  8.31$\pm$0.02 \\
NGC 5194  &      $203.00\pm14.00^{b}$  & 18.00$\pm$0.38 &  28.20$\pm$1.31 &   1.62$\pm$0.06 &   1.15$\pm$0.07 	&  20.70$\pm$0.33 &  31.80$\pm$0.33  &  9.05$\pm$8.79 &    7.59$\pm$7.33  &  8.55$\pm$0.01 \\
NGC 5033  &       $40.78\pm2.85^{a}$   & 19.20$\pm$0.27 &  20.10$\pm$0.31 &   1.59$\pm$0.04 &   1.54$\pm$0.04 	&  27.30$\pm$0.23 &  33.00$\pm$0.23  &  9.42$\pm$9.10 &    9.26$\pm$8.94  &  8.56$\pm$0.02 \\
NGC 1377  &        $1.22\pm0.08^{a}$   & 20.70$\pm$0.38 &  31.80$\pm$1.22 &   1.48$\pm$0.04 &   1.03$\pm$0.05 	&  17.70$\pm$0.32 &  20.70$\pm$0.32  &  8.47$\pm$8.57 &    7.17$\pm$7.27  &  8.29$\pm$0.10 \\
NGC 1482  &       $15.40\pm1.00^{a}$   & 27.30$\pm$1.10 &  33.00$\pm$1.79 &   1.17$\pm$0.06 &   0.99$\pm$0.07 	&  33.90$\pm$0.92 &  39.60$\pm$0.92  &  8.77$\pm$8.43 &    8.25$\pm$7.91  &  8.11$\pm$0.13 \\
NGC 1705  &        $0.85\pm0.13^{c}$   & 17.70$\pm$0.61 &  20.70$\pm$0.97 &   1.65$\pm$0.10 &   1.47$\pm$0.10 	&  17.10$\pm$0.53 &  21.30$\pm$0.53  &  4.36$\pm$3.99 &    3.78$\pm$3.52  &  8.27$\pm$0.06 \\
NGC 3773  &        $0.94\pm0.07^{a}$   & 33.90$\pm$0.38 &  39.60$\pm$3.34 &   0.97$\pm$0.08 &   0.78$\pm$0.09 	&  17.10$\pm$0.32 &  21.30$\pm$0.32  &  5.09$\pm$4.96 &    4.67$\pm$4.54  &  8.43$\pm$0.03 \\
NGC 5866  &        $7.57\pm0.53^{a}$   & 17.10$\pm$0.21 &  21.30$\pm$0.73 &   1.68$\pm$0.05 &   1.49$\pm$0.05 	&  17.10$\pm$0.18 &  21.30$\pm$0.18  &  7.42$\pm$7.06 &    6.66$\pm$6.30  &  8.47$\pm$0.10 \\
\hline
 \end{tabular}
\end{table*}

\subsection{The dust spatial distribution vs. optical stellar continuum emission extents}\label{subsec:Rdust_vs_Rdisk_RHalpha}

\begin{figure}
  \includegraphics[scale=0.40]{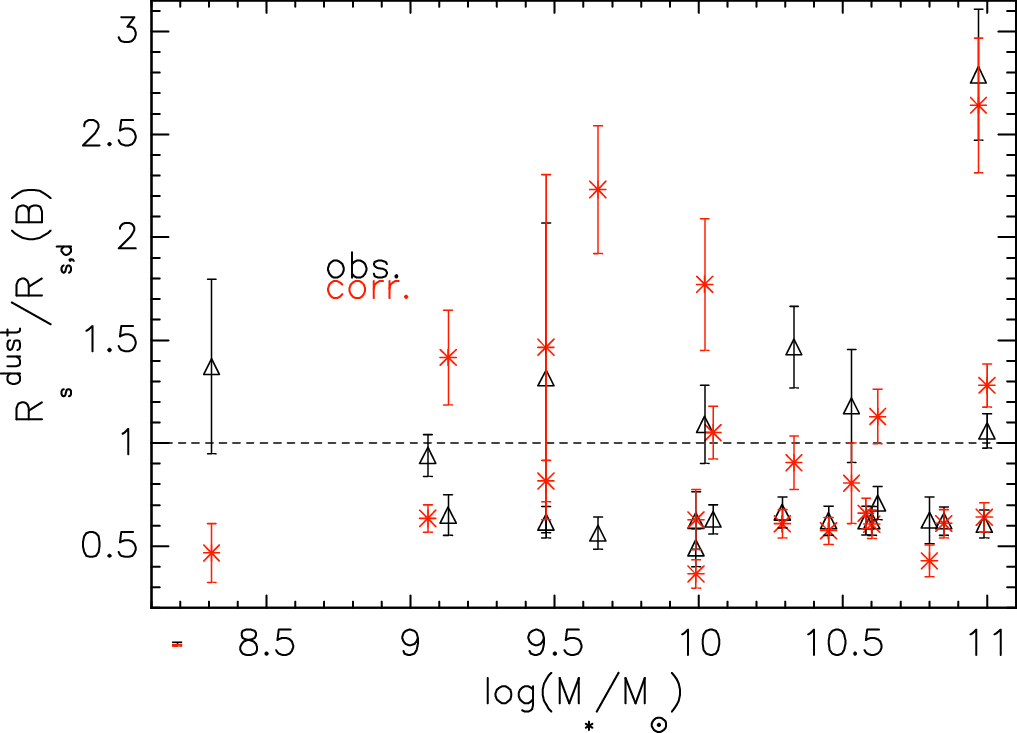}
    \includegraphics[scale=0.40]{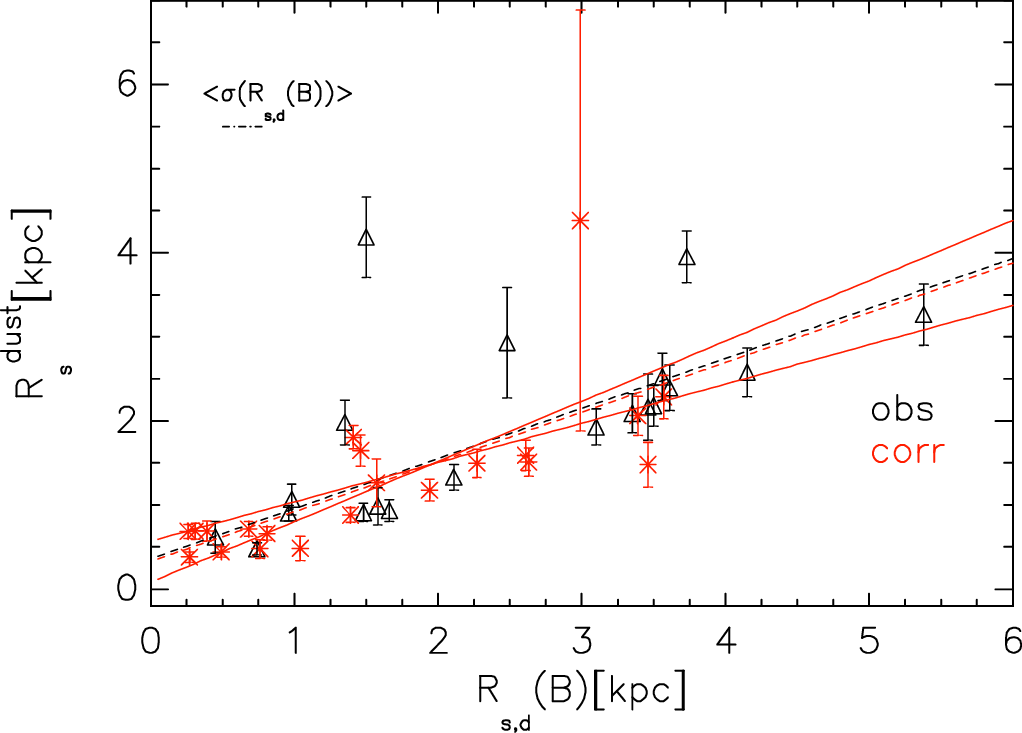}
  \caption{\label{fig:Rs_ratio_Mstar} \textit{Upper panel} The ratio between the observed (black triangles) / intrinsic (red stars) scale-lengths of the
  dust discs and the ones of the stellar emission discs (seen in optical B band), as a function of the stellar mass.
  \textit{Lower panel} The same scale-lengths, plotted agains each other. The black and red dotted line show a linear-regression best-fit of the observed and corrected
  values. The rest of symbols, colors and lines have the same meaning as those in Fig.~\ref{fig:Tdust_compare_plots_beta_free_vs_fixed}.}
\end{figure}

We investigated here how does the dust spatial distribution compares with the optical continuum, corresponding to the stellar emission. We have already seen in
Paper III that the star-formation disc is slightly more extended than the stellar emission one (see Fig. 8). As both components are considered to have an exponential disc
type of distribution, we compared the scale-lengths of the dust and stellar discs. We plotted their ratio, $R_{s}^{dust}/R_{s,d}(B)$, as a function of the stellar mass, in the
upper panel of Fig.~\ref{fig:Rs_ratio_Mstar}, for the observed (black) and corrected values (red). The B band stellar disc scale-lengths, $R_{s,d}(B)$ were determined in Paper I.
One can see that on average, the dust disc scale-lengths are slightly smaller than the optical stellar disc ones, and their is a very mild increase of their ratio with
$M_{\star}$. We derived the average observed ratios of $0.84\pm0.15 / 0.94\pm0.14$ (with / without outliers), and $0.78\pm0.16 / 0.80\pm0.15$ respectively (with / without
outliers), for the intrinsic ones. We can see that for some cases the ratios are up to 1.5 and beyond, which means that the exponential dust discs are considerably more extended
than their stellar ones. \cite{MunM09} fitted the dust surface density profiles of SINGS galaxies with an exponential function and found an average factor of 1.1 between the
derived dust disc scale-lengths and the stellar emission ones derived from data at 3.6$\mu m$.\\
\cite{Casa17} determined likewise an average ratio of 1.1 for the scale-lengths at 250$\mu m$ and 3.6$\mu m$, while \cite{Hunt15}, by analysing the whole KINGFISH sample through
exponential fits of the radial profiles at the same 3.6$\mu m$ and 250$\mu m$ wavelengths, determined a mean value of $0.96\pm0.36$ for the ratio of exponential scale-lengths
of the cool dust and stars distributions. Therefore, our results are in agreement (within errors) with these studies, although a direct comparison is not possible due to 
different wavelengths considered within each study. Nevertheless, our ratios (which consider the B band stellar emission scale -lengths) can be more relevant in this context,
since dust emission is connected in a higher proportion to the young stellar population.

We also plotted in the lower panel of the same figure the two scale-lengths against each other. A linear monotonic increase can be noticed, which means that galaxies with
a larger extent of the stellar disc also have a larger extent of the dust disc. We determined a slope of $\alpha=0.59\pm0.09$ for the intrinsic (dust-corrected) ratios, 
consistent within errors with the value of 0.69 found by \cite{Qin24} (from their model) but slightly lower than the ones found by \cite{Mos22} - $0.81\pm0.31$, for the scalengths at 100 and 3.4$\mu m$), and \cite{Casa17}, who found $0.83\pm0.05$ for the pair of scale-lengths at 100 and 3.6$\mu m$. The derived correlation coefficient
$r_{R_{d}^{dust},R_{s}(B)}=0.55$ is close to the value of 0.60 derived in \cite{Mos22} for their scale-lengths. Of course, as already mentioned here, we need to remember that
the ratio of dust and stellar disc scale-lengths depends on the wavelengths, but also on the redshift (if we consider the inside-out galaxy evolution scenario the stellar disc
extent will increase as the redshift decreases, whereas the precise evolution of the dust disc size is more complicated to predict), and so a direct and meaningful comparison with the previously
mentioned studies (and others), is not straightforward. Moreover, the stellar disc scale-length is affected mainly by dust effects, and to a lesser extent by projection and
decomposition effects (\citealt{Pas13a, Pas13b}). Therefore, a more meaningful comparison would be between the intrinsic (dust-corrected) ratios - the red values, and 
similar intrinsic ratios from other studies, where available.

\section{Discussion}\label{sec:discussion}

In this section we are coming back to a few issues noticed when analysing the main results, evaluate the potential sources of systematic errors and other uncertainties,
and discuss the limitations of the method.

\subsection{Potential sources of systematic errors}

The most important source of systematic errors in this study is the consideration of a single-temperature MBB function to characterise the cold dust emission in the NIR-submm range,
that does not take into account a warmer dust component. The end result is a mean dust temperature. However, this approach has been largely used in the literature, producing
noteworthy results. One way to ameliorate this is the consideration of two modified blackbody functions to characterise the cold and warm dust emission (each component with its
own temperature and $\beta_{d}$) or a broken MBB (BMBB, with a broken emissivity law; \citealt{Gor14}) - which allows for a variation in the wavelength dependence of the dust 
emissivity law (therefore having two $\beta_{d}$ values for shorter and longer wavelengths, and a break wavelength as a parameter), to account for a sub-mm excess. These 
approaches have been attempted in \cite{Lamp19} on a sample of 192 low-redshift galaxies and revealed very small changes in the determined dust masses between each of these
three situations. The dust temperatures derived for the BMBB are slightly lower than for the single MBB case, while $\beta_{d}$ of the cold dust component is  slightly larger.
In the case of two MBB, the authors found a smaller $T_{d}$ than for the single MBB case, with dust emissivity indices similar for both. Having already the dust masses previously
determined (a single set of measurements, for the cold dust component), it is not straightforward to check these two scenarios for our sample, particularly the one with two MBB 
(in which case we would need two sets of dust masses derived apriori, for the cold and warm components). In another study, using their analytical model to create dust SEDs and
testing it on high-redshift galaxies, \cite{SomAlg25} found that considering multi-temperature dust can determine significantly underestimated dust masses and systematically 
flatter $\beta_{d}$ values. \\
Related to the systematic errors caused by using a single-temperature MBB is the consideration of the cold dust SED peak at a fixed $\lambda_{peak}=250\mu m$ for the whole
sample. Despite the fact that all galaxies analysed here are at very low-redshifts, some differences may exist in the values of $\lambda_{peak}$ between galaxies which can 
introduce a systematic bias. However, we appreciate this bias to be within the error limits of the derived dust temperatures and emissivity indices.\\
Another way to improve this approach and better constrain $\beta_{d}$ would be to consider the Rayleigh-Jeans side of the SED (fluxes in the sub-mm range)
and use the slope of this part of the SED to derive this parameter (e.g. \citealt{Bendo24}). This would of course require observational data in that wavelength range, that
could be available from surveys such as ALMA or the proposed Atacama Large Aperture Submillimeter Telescope (AtLAST, \citealt{Liu25}). \\
Another source of uncertainty would be the consideration of a fixed $\kappa_{\nu_{0}}$ characteristic for a dust model with a fixed value for $\beta_{d}$, as the dust mass 
absorption coefficient is not very well constrained and it is not certain how it varies with radius. However, we have shown in Sect.~\ref{subsec:dust_temp_beta} that this
effect is cancelled here and this should not bias the derived ($T_{d}, \beta_{d}$) values. As we have used dust masses derived separately from this study, to determine dust
temperatures and emissivity indices, a systematic bias (that cannot be overcome at this point) can be introduced by the choice of the \cite{Draine03} dust model, its 
characteristics being encapsulated in the $K(B)$ constant.\\
Connected with this is the fact that we considered $\kappa_{\nu_{0}}$ at a reference wavelength $\lambda_{0}=100\mu m$. As in other studies various reference wavelengths
were used (e.g. 250$\mu m$, 350$\mu m$, 850$\mu m$ or others), a systematic bias can be introduced by this choice. But that would be true even with a different choice of
$\lambda_{0}$, and of course some small differences may arise in the best-fit values of $(T_{d}, \beta_{d})$ as a result of this. However, the variation of $(T_{d}, \beta_{d})$ 
with $\lambda_{0}$ is beyond the scope of this paper.

\subsection{Limitations of the method and range of applicability}
    
The limitations of the proposed method to derive $(T_{d}, \beta_{d})$ values and its range of applicability are tightly connected to the range of applicability of the
relations used to determine the dust massses (and dust optical depths needed for $M_{d}$) in Paper I - Eqs.~\ref{eq:Grootes}\&\ref{eq:Mdust_tauB}. These in turn depend on
the range of applicability of the fixed large-scale dust-star geometry (dust and stars distributed in exponential discs) as calibrated in the \cite{Pop11} model to the 
range of galaxy types, morphologies (same as for the dust and inclination effects numerical corrections) and stellar mass surface densities, $8.0 \leq log(\mu_{\ast}) \leq 11.0$
(therefore intermediate mass galaxies), one has to analyse with this method.\\
Another limitation would be the lack of a 250$\mu m$ flux / luminosity for certain galaxies. However, one can use in that case the 160$\mu m$ flux or luminosity (if 
available), especially for higher redshift galaxies. We tested this for our nearby galaxies to check if / how the results would change in this case. We noticed that 
differences arise in the $T_{d}$ and $\beta_{d}$ values, but the trends seen in the analysed relations are conserved, with some small differences in their characteristic
parameters.

\section{Summary and conclusions}\label{sec:conclusions}

In this paper we have presented a simple method to derive dust temperatures and dust emissivity indices, based on previous results - the dust masses derived (independent of
this study) in Paper I and previous prescriptions. For consistency, and as a case study for validation of the method, we have used the sample of spiral and lenticular 
galaxies from the KINGFISH survey, for which we have determined the dust masses and other quantities.\\
A modified blackbody function was considered to characterise the cold dust emission SED in the NIR-submm wavelength range, while only the flux / luminosity at 250$\mu m$,
characteristic for the peak of the cold dust emission, was required for the presented approach. Thus, a time consuming whole SED fit was not necessary. We derived dust 
temperatures and corresponding dust emissivity indices for the same galaxies analysed in Papers I and III, through a grid-type procedure. Two scenarios were considered - 
one where $\beta_{d}$ was left free, and another where $\beta_{d}$ was fixed to a value commonly used in other works, $\beta_{d}=2$. We then quantified the $T_{d}-\beta_{d}$
evolution (and its potential degeneracy), and investigated new or confirmed relations between $\Sigma_{d}, \beta_{d}$, metallicity, SFR, and their implications for ISM 
evolution and star-formation.\\
Our main results are as follows:
\begin{itemize}
\item the hyperbolic shape of the $T_{d}-\beta_{d}$ anti-correlation is again found, despite using dust masses that are not an output of a full cold dust SED fit but are 
derived through another independent method, and without the noise associated with all the observed NIR-FIR-submm fluxes; this fact suggests that this anti-correlation may
have, at least partially, a physical origin;
\item the means for observed and intrinsic $T_{d}$ and $\beta_{d}$ (when $\beta_{d}$ is left free) and their range of values are consistent with many other similar studies;
when $\beta_{d}=2$, both the observed and corrected average dust temperatures are \textit{systematically higher};
\item the mild anti-correlations in $T_{d}-12+[O/H]$ and $T_{d}-\Sigma_{d}$ show that metal-rich galaxies have colder dust on average, while colder dust is more concentrated
in the disc of the galaxies, if $\beta_{d}$ is left free; when $\beta_{d}=2$, the opposite happens for the latter relation;
\item SFR, sSFR and $\Sigma_{SFR}$ are not conclusive tracers of dust temperature, in both scenarios;
\item metal-rich galaxies tend to have dust with higher SFR and $\beta_{d}$, and therefore a grain composition made predominantly from crystalline silicates;
\item $\Sigma_{d}$ correlates strongest with $M_{\ast}$, but this a consequence of the fundamental $M_{d}-M_{\ast}$ relation; \textit{new, mild correlation $\Sigma_{d}-\beta_{d}$}
is observed;
\item dust is more compactly distributed in galaxies with lower sSFR, and therefore higher stellar mass and higher dust attenuation;
\item a linear monotonic increase of the sizes of stellar discs with the dust disc ones is noticed, with a slope $\alpha=0.59\pm0.09$, consistent with other results;
\item on average, the dust disc scalele-lengths are comparable with the scale-lengths of the optical stellar discs, slightly smaller, with some cases where these are considerably
higher.
\end{itemize}
While this was a case study, where we have used a well studied small sample of representative galaxies in the nearby Universe, the same simple procedure can be applied to other
suitable, larger samples of low to mid-redshift normal galaxies, using the suite of prescriptions presented in Paper I (needed for dust opacity and dust mass calculation),
III (for SFR and related parameters) and in this paper (to derive dust temperature and dust emissivity index), for larger scale studies of star-formation and ISM evolution.

\section*{Acknowledgements}
This research made use of the NASA/IPAC Extragalactic Database (NED), which is operated by the Jet Propulsion Laboratory, California Institute of Technology, under contract
with the National Aeronautics and Space Administration.\\
This research was supported by the Romanian Ministry of Research, Innovation and Digitalization under the Romanian National Core Program
LAPLAS VII - contract no. 30N/2023

\section*{Data Availability}
The data underlying this article are available in the article.\\


\bsp	
\label{lastpage}
\end{document}